\def\rocket{\epsfysize=6 cm
 \epsfbox{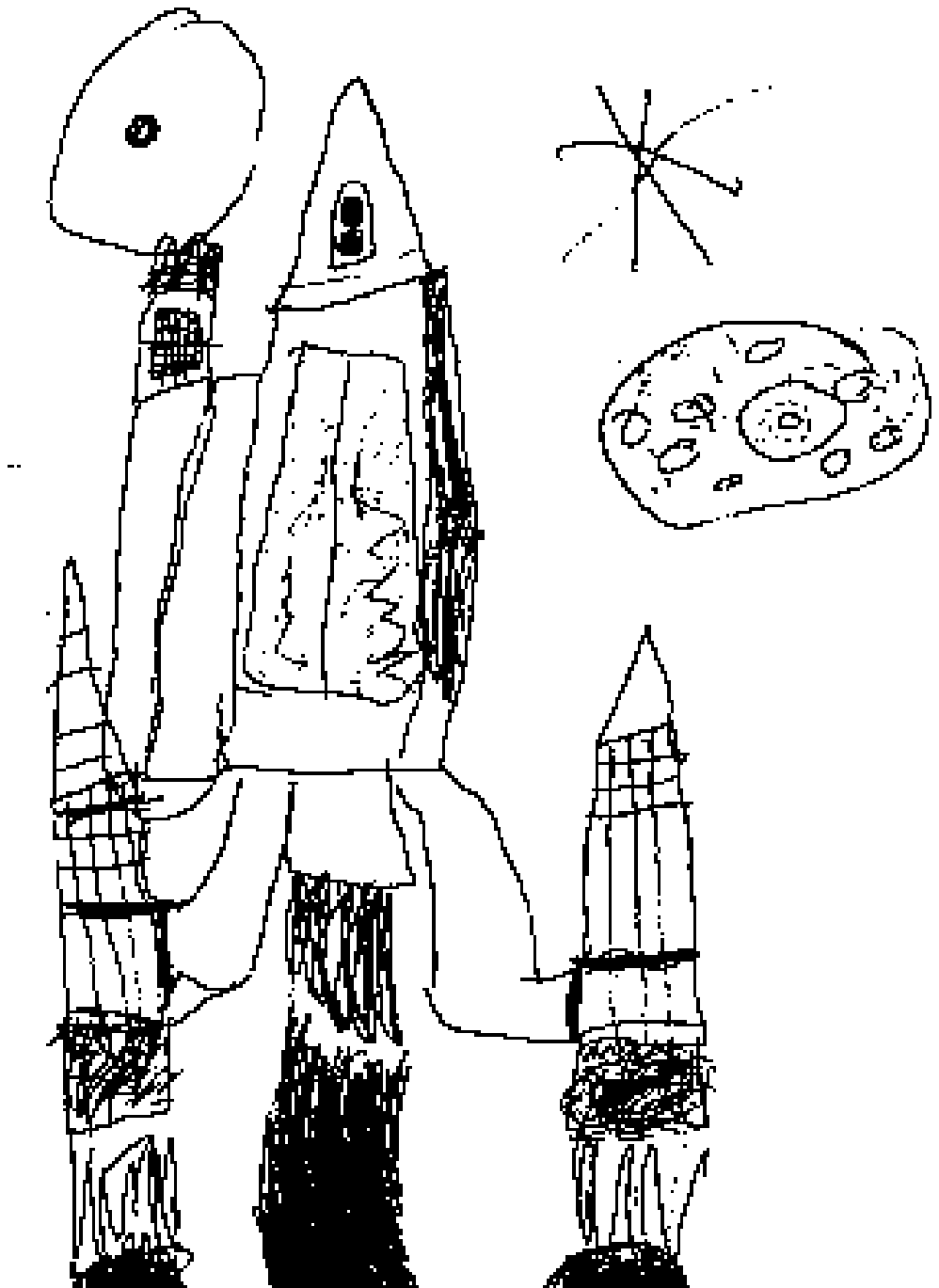}}
\def\thorne{\epsfysize=6 cm
 \epsfbox{thorne.EPSF}}

\def\einstein{\epsfysize=6 cm
 \epsfbox{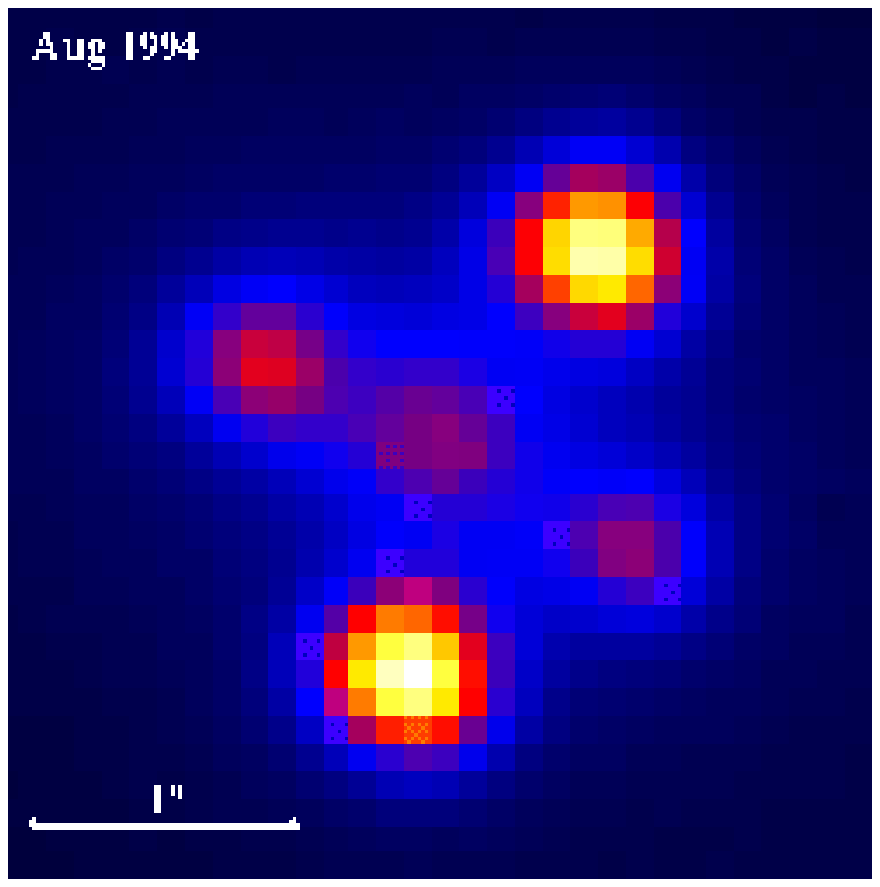}}

\def\suivi2{\epsfysize=5.0 cm
 \epsfbox{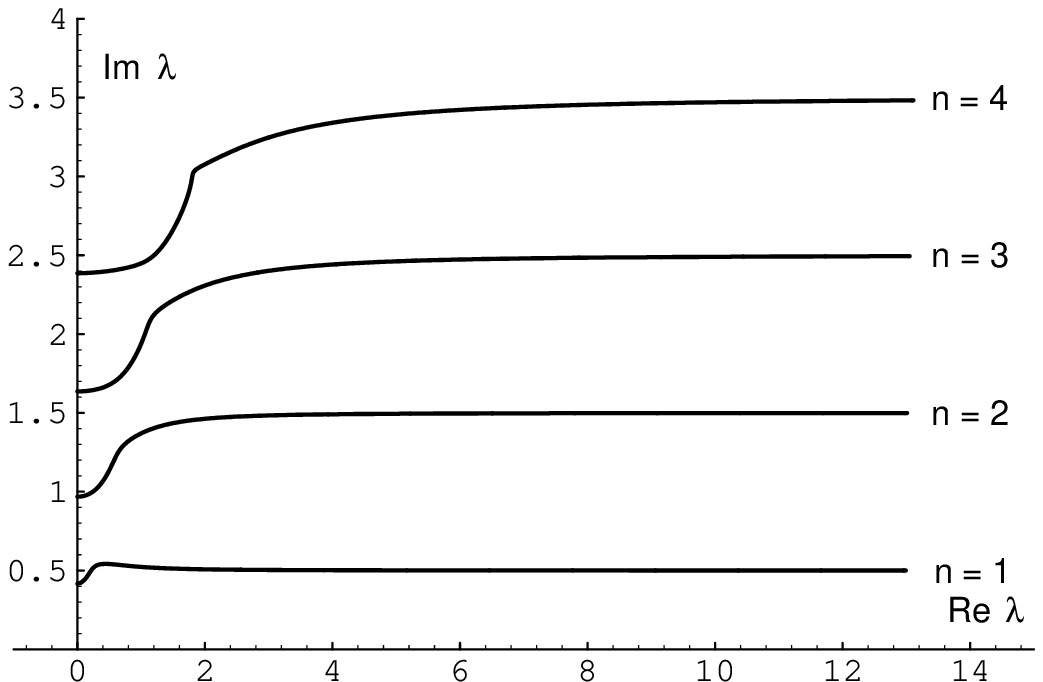}}
\def\contour1{\epsfysize=5 cm
 \epsfbox{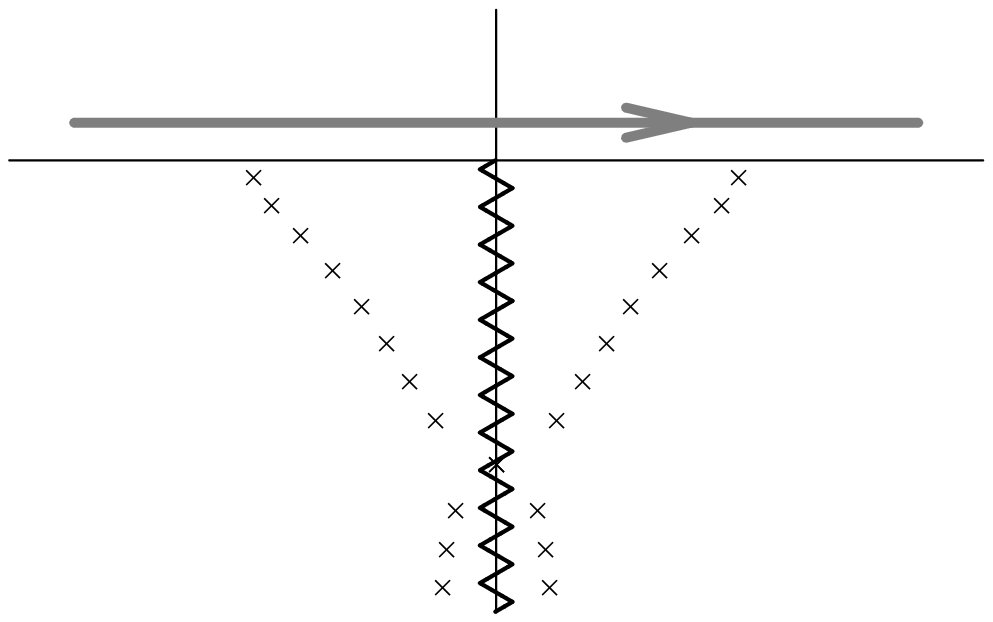}}
\def\leaver{\epsfysize=5 cm
 \epsfbox{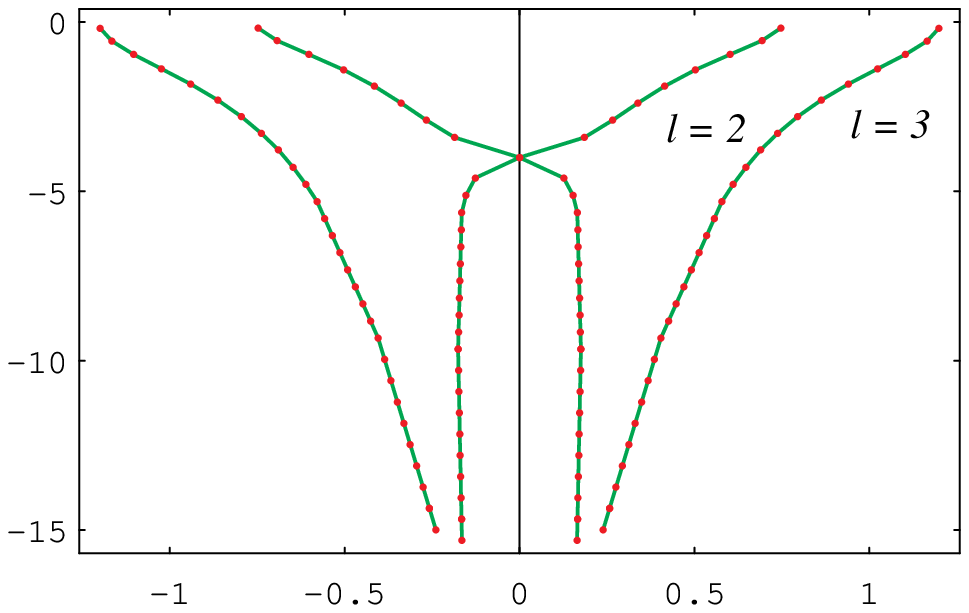}}
\def\nulgeos{
\epsfysize=6 cm
 \epsfbox{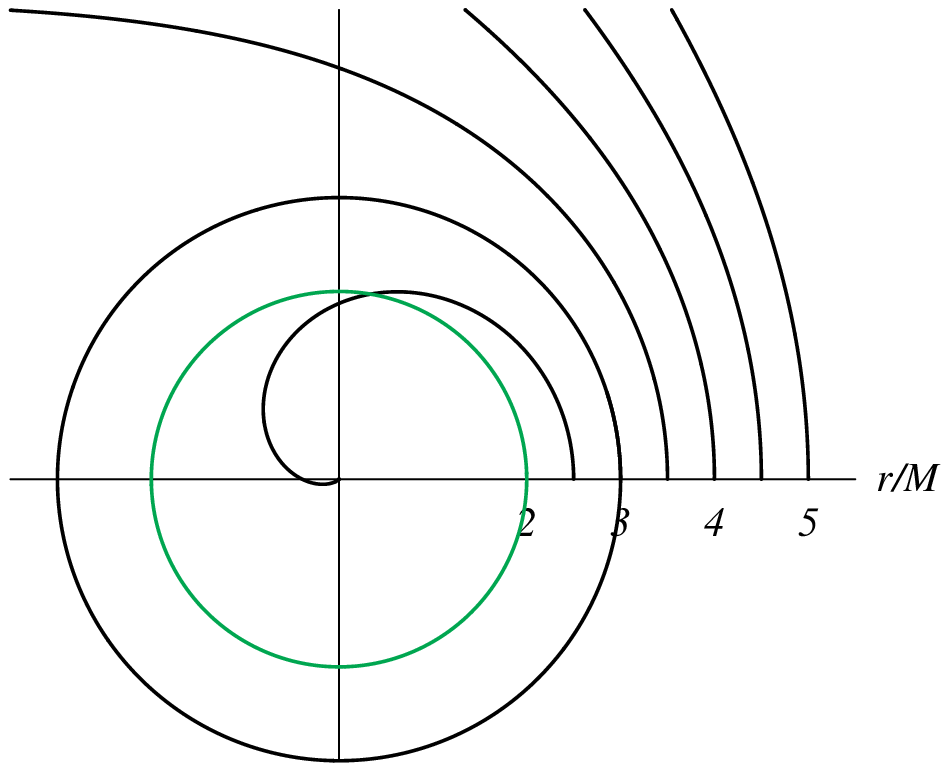}} 
\def\hump2{ \epsfysize=5 cm
 \epsfbox{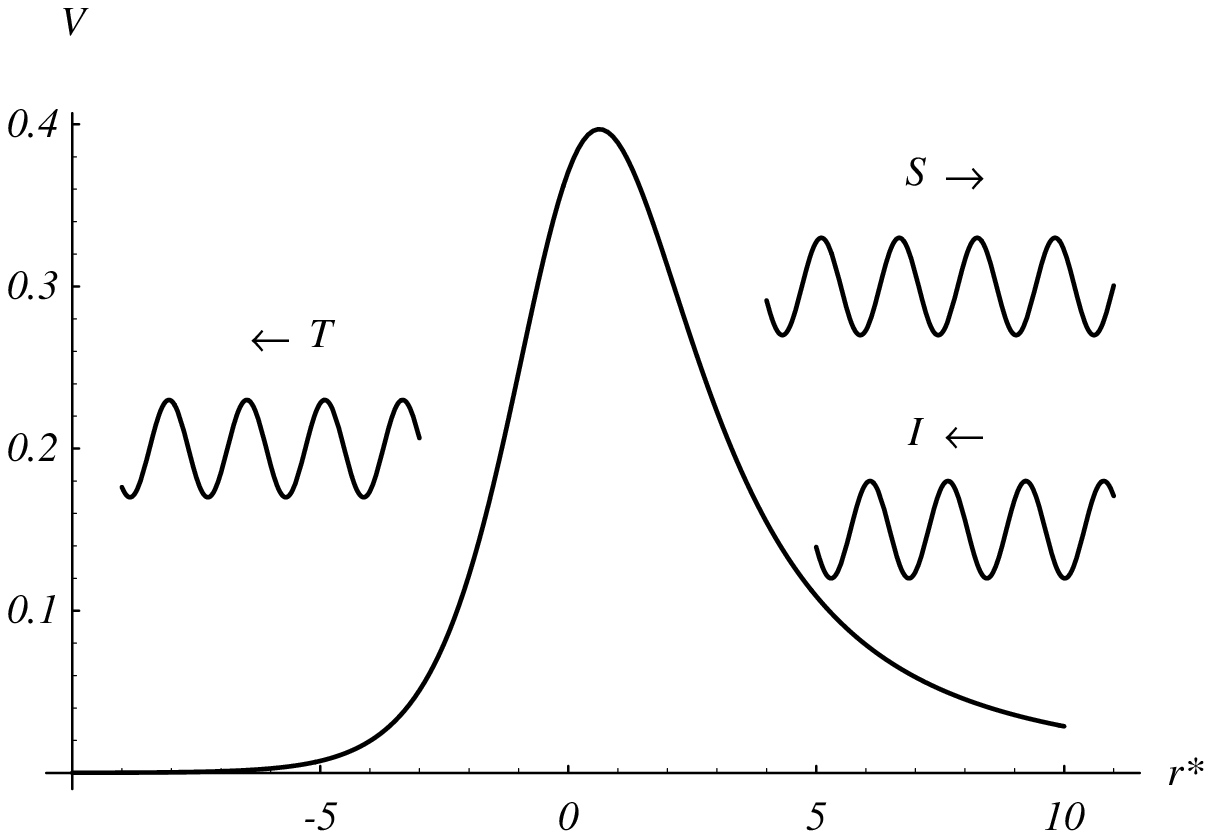}}
\def\glens{ \epsfysize=5 cm
 \epsfbox{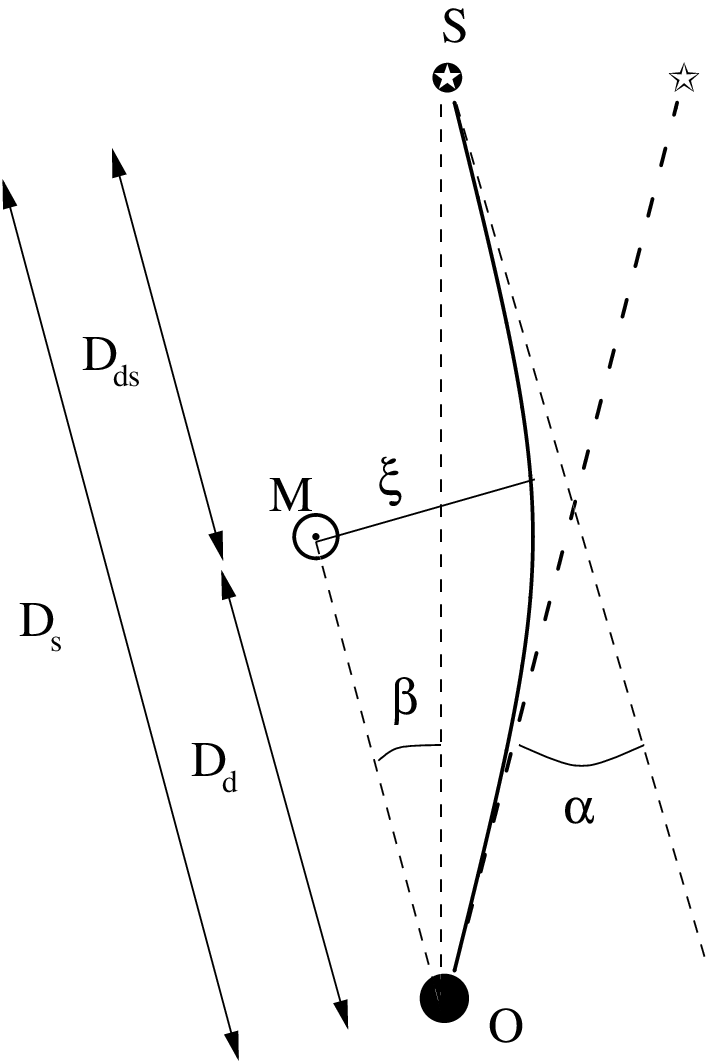}}
\def\inup{ \epsfysize=5 cm
 \epsfbox{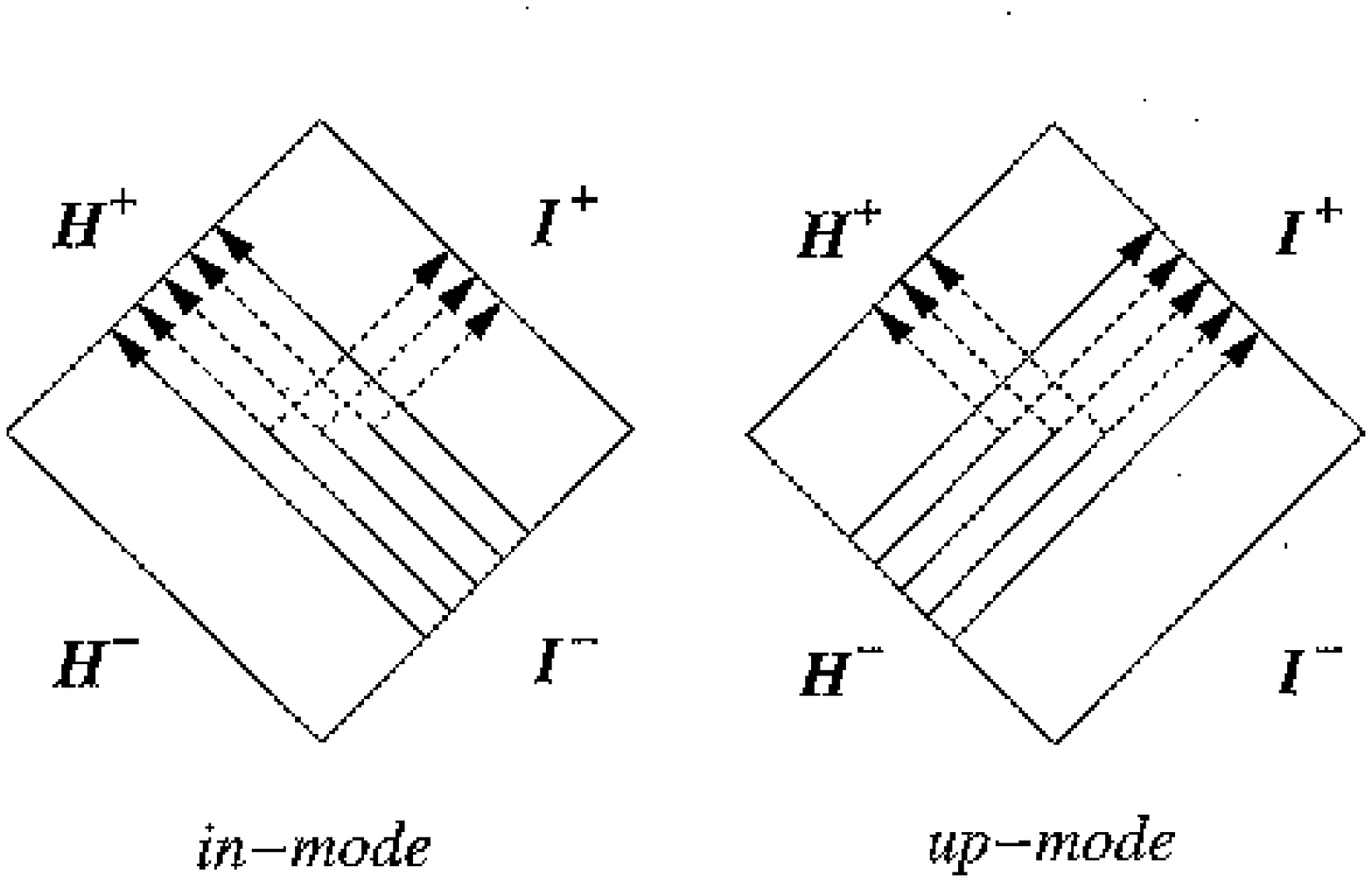}}
\def\deflect{ \epsfysize=5 cm
 \epsfbox{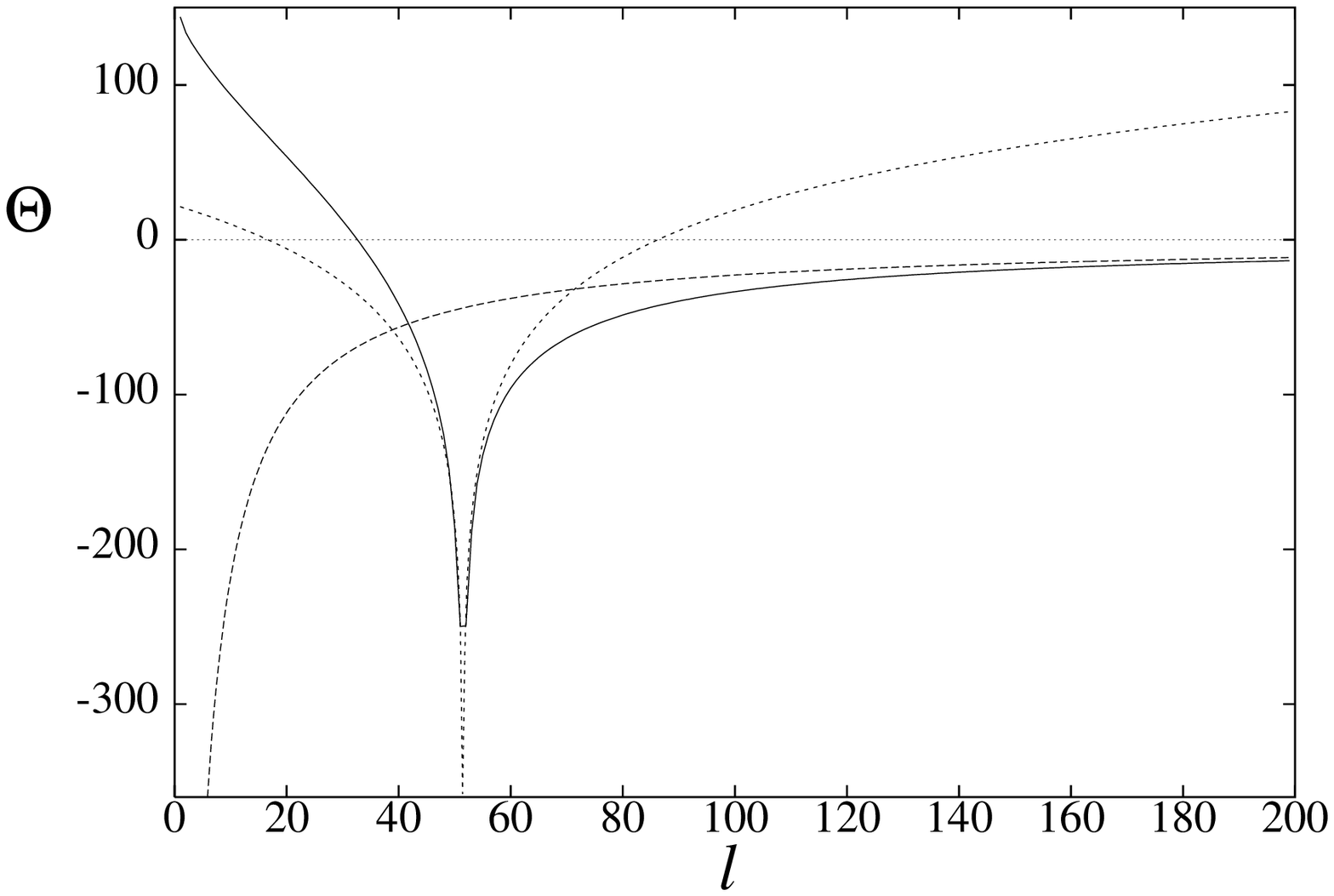}}
\def\cross{ \epsfysize=5 cm
 \epsfbox{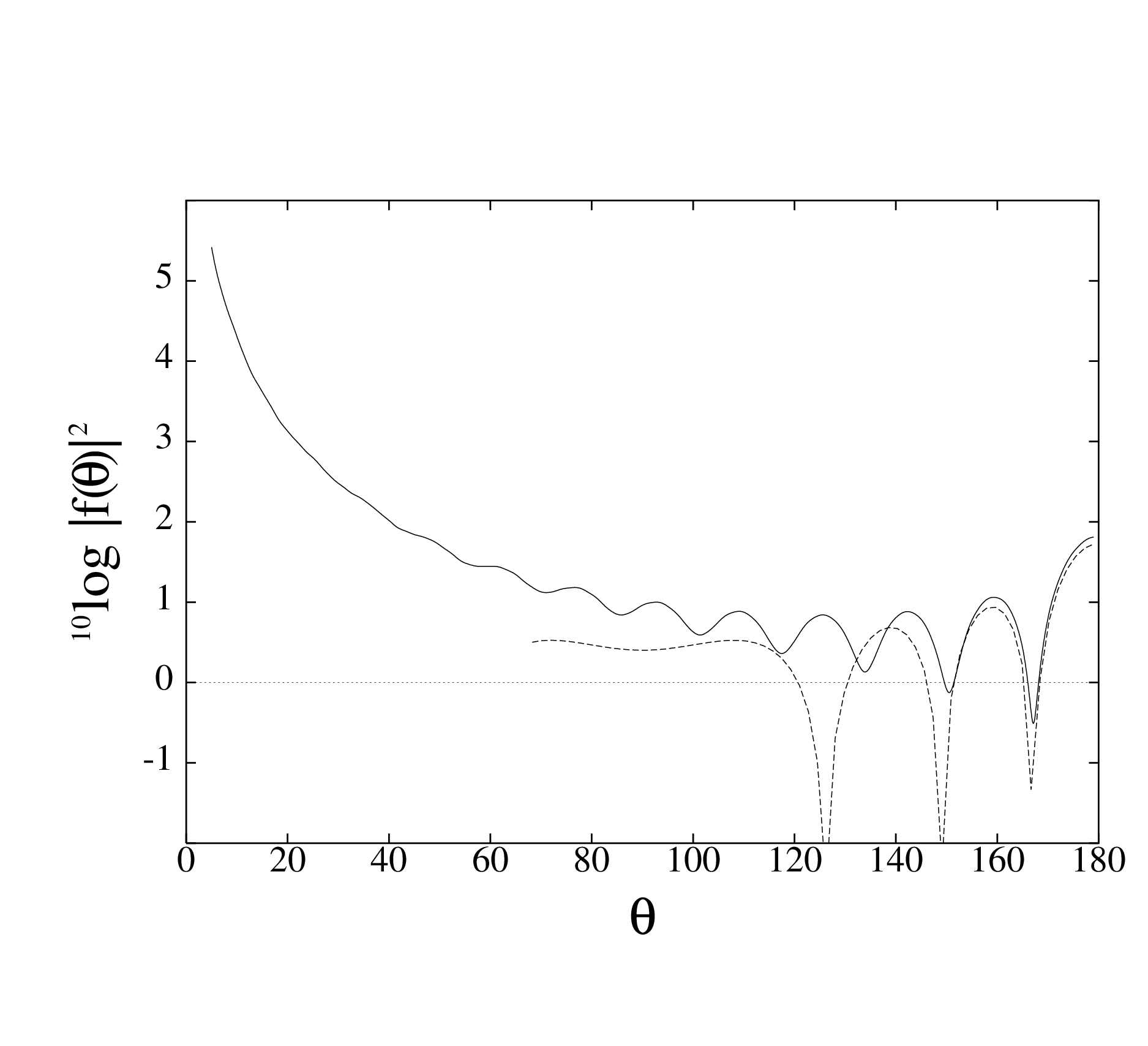}}
\def\vishu{ \epsfysize=8 cm
 \epsfbox{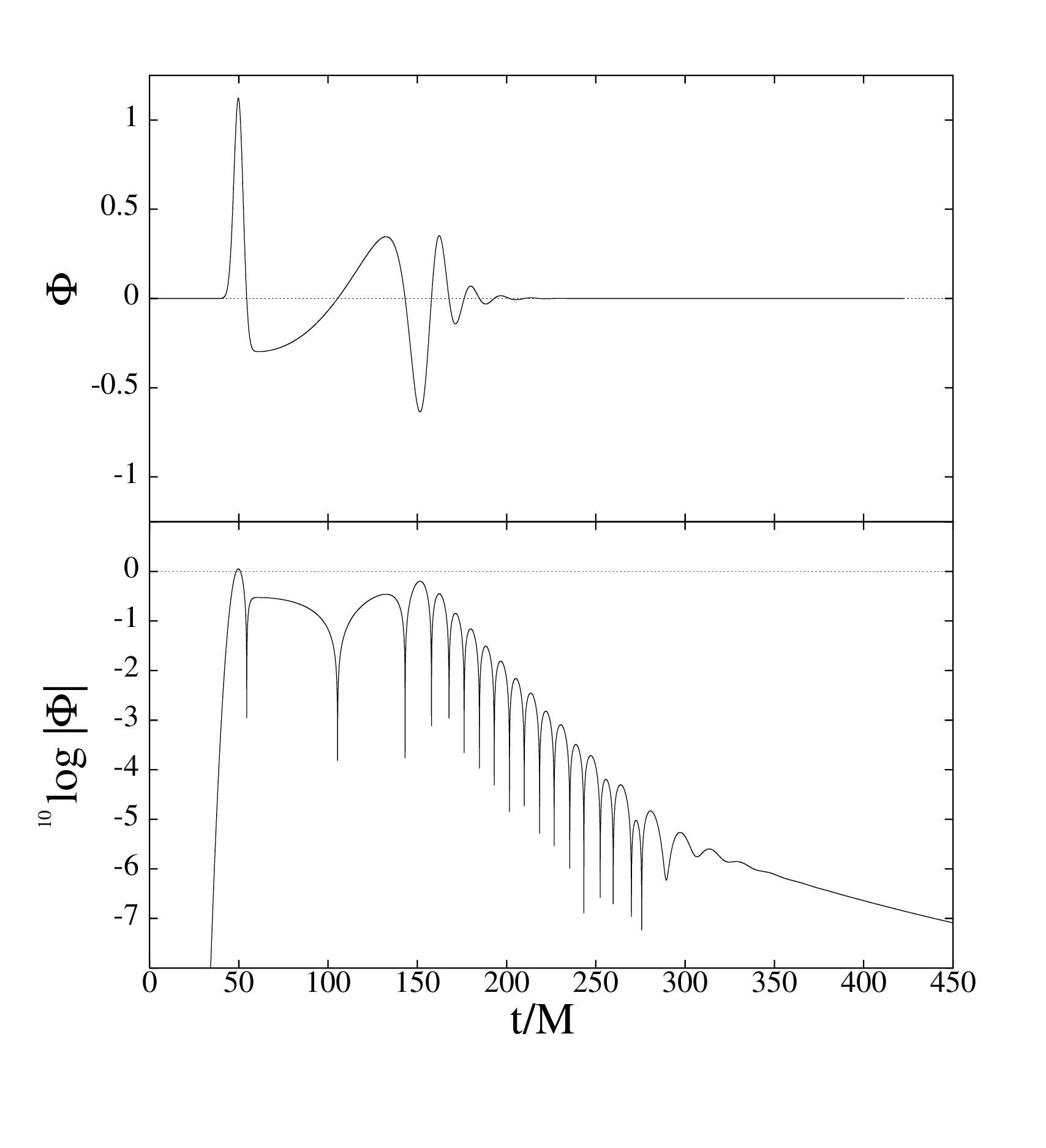}}
\def\anglesp{\epsfysize=5 cm
 \epsfbox{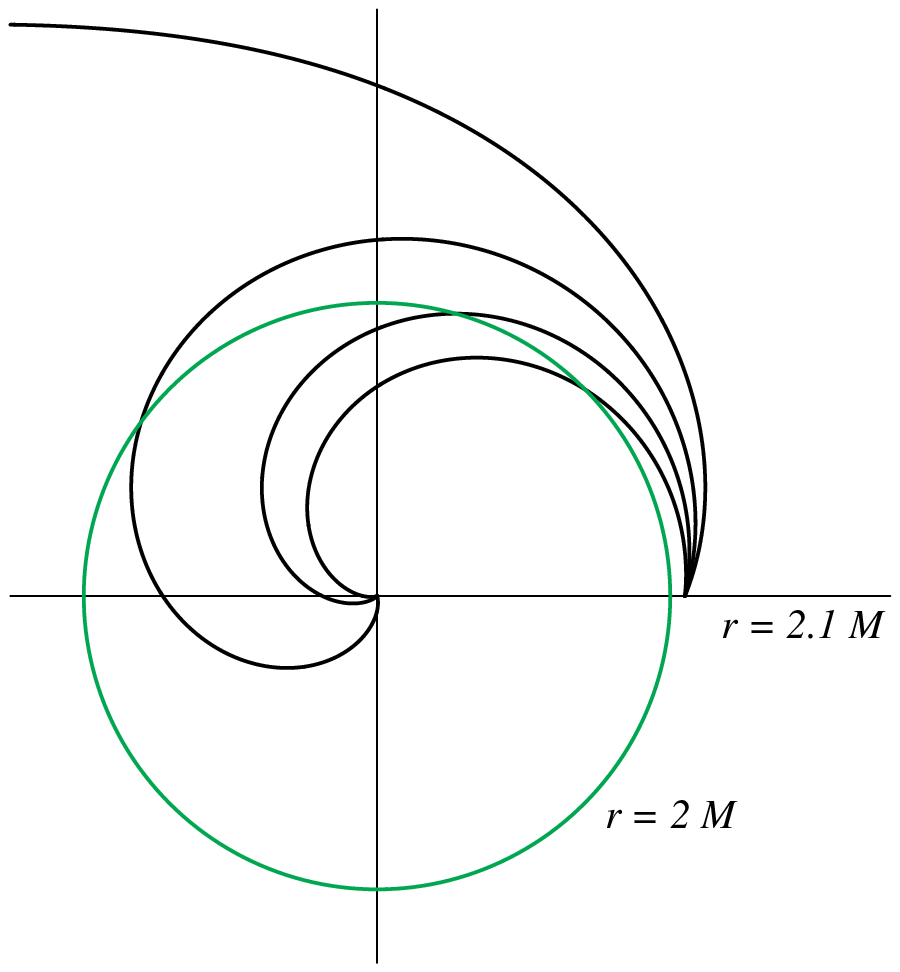}}

\def\scat{\epsfysize=5 cm
 \epsfbox{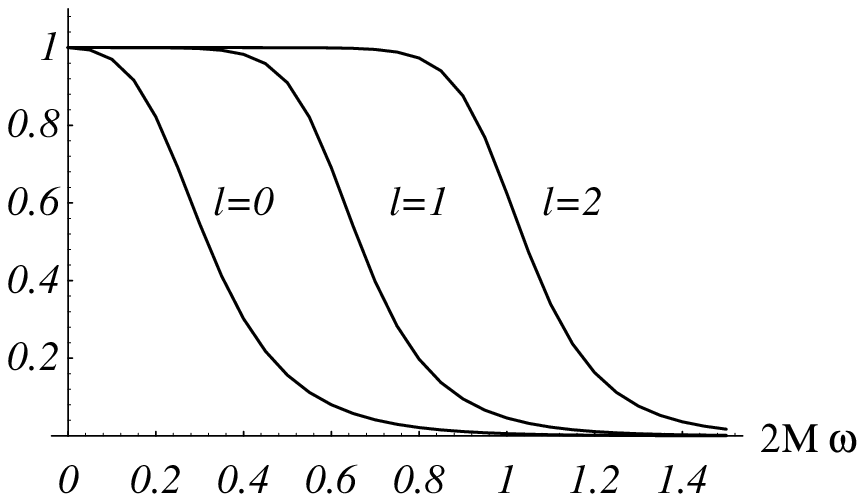}}
\def\rcross{\epsfysize=5 cm
 \epsfbox{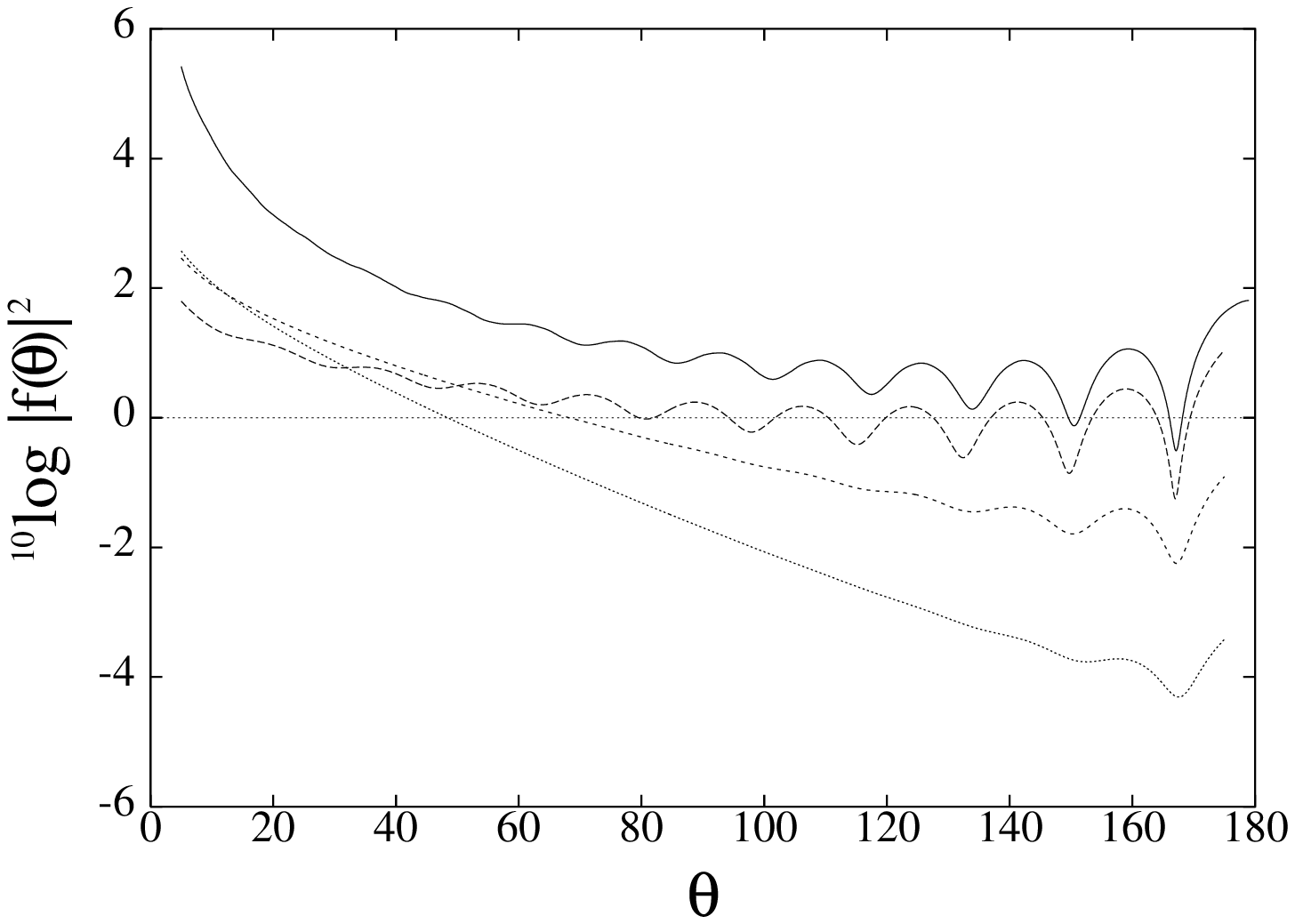}}
\def\vkerr{\epsfysize=5 cm
 \epsfbox{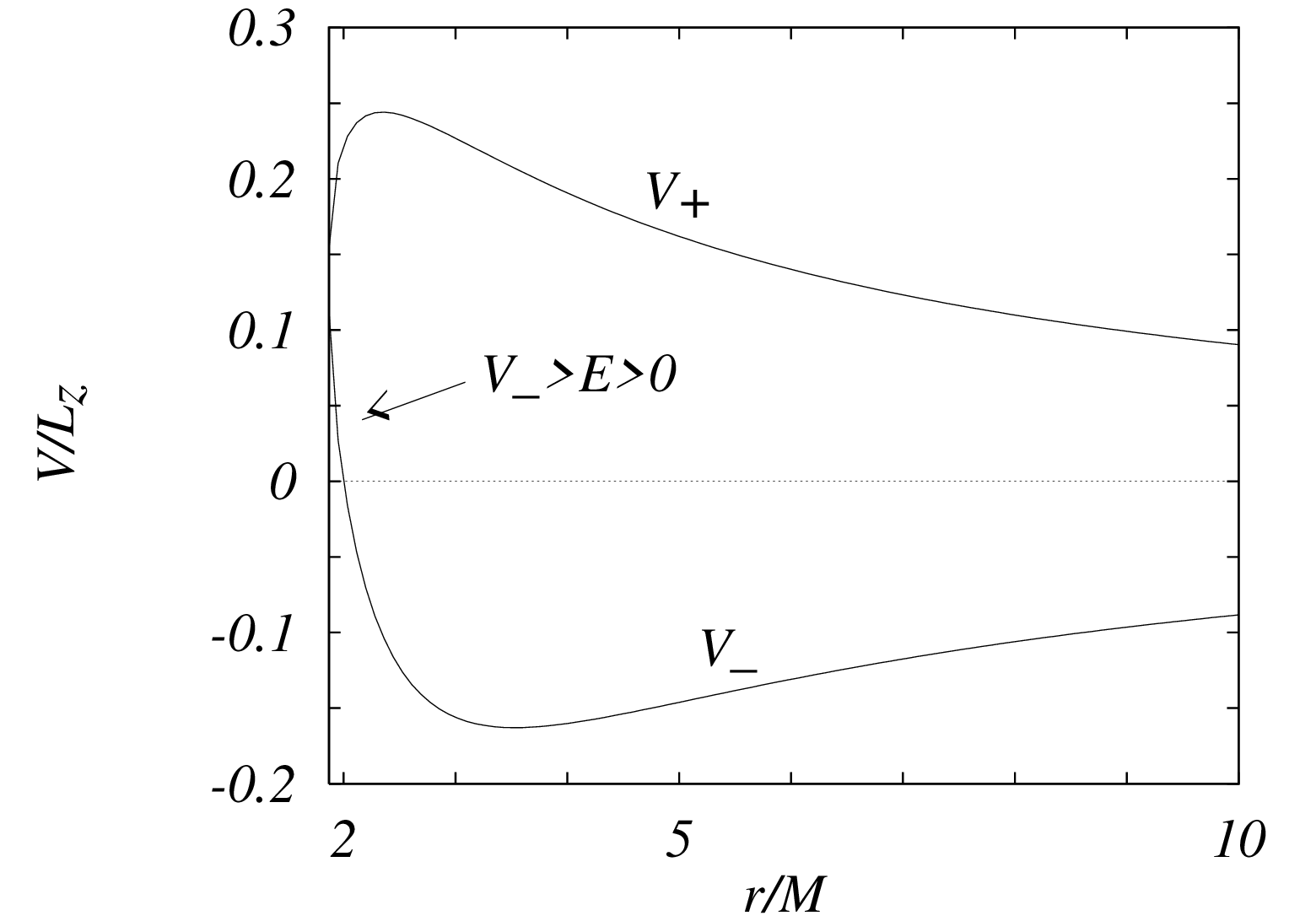}}
\def\tkev{\epsfysize=5 cm
 \epsfbox{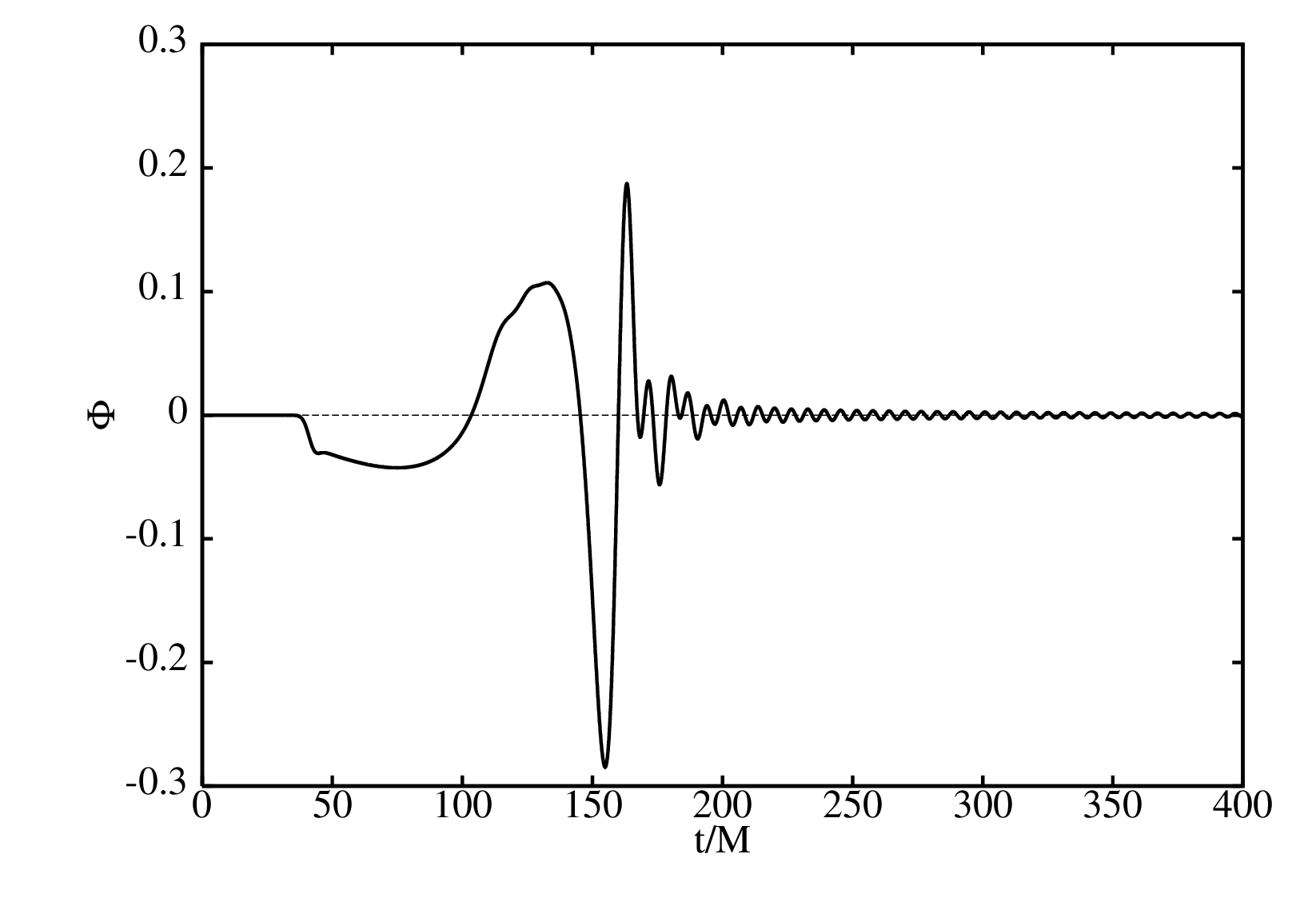}}
\def\srad{
\epsfysize=8cm
\epsfbox{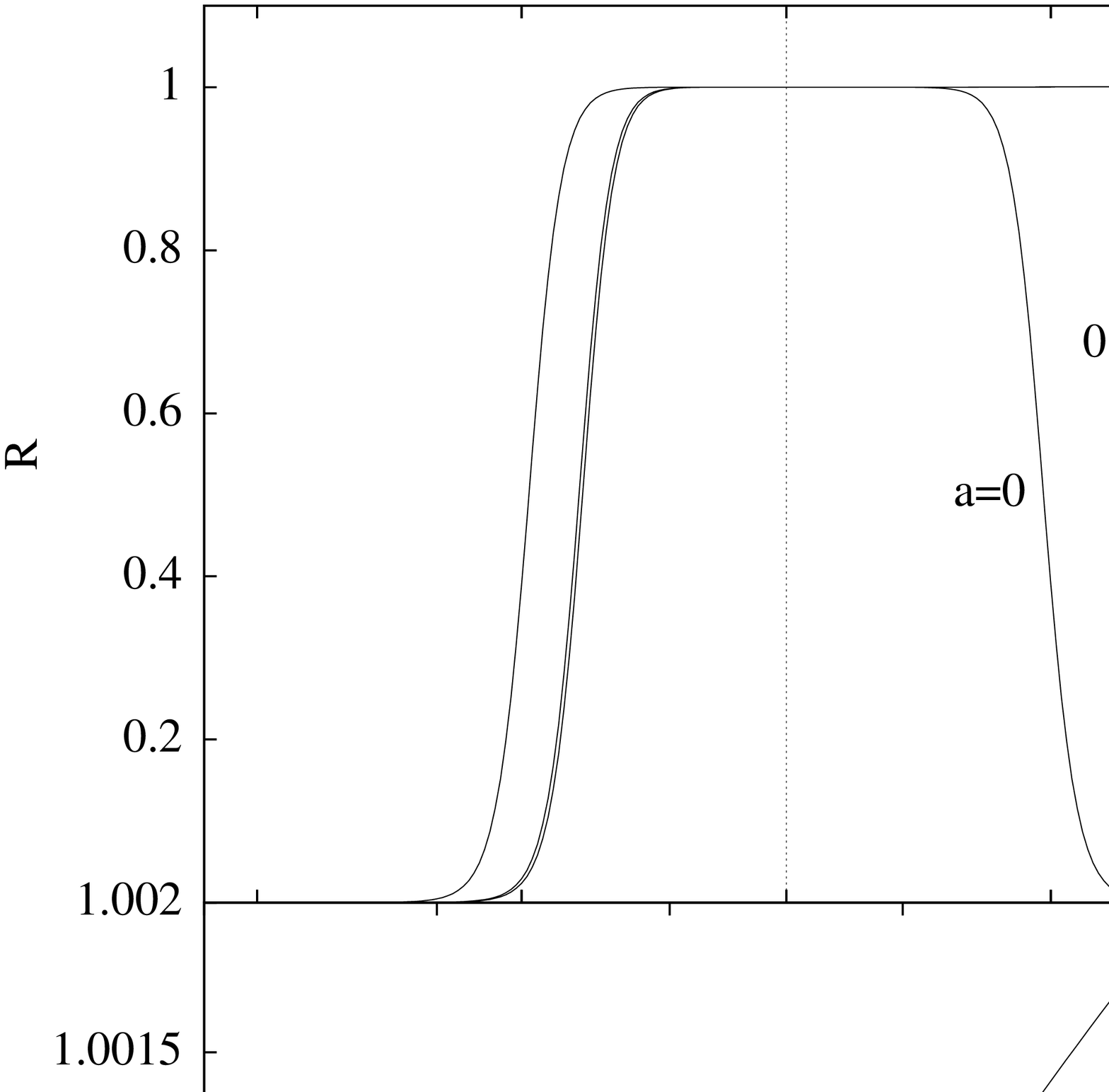}}

\def\cont2{\epsfysize=5 cm \epsfbox{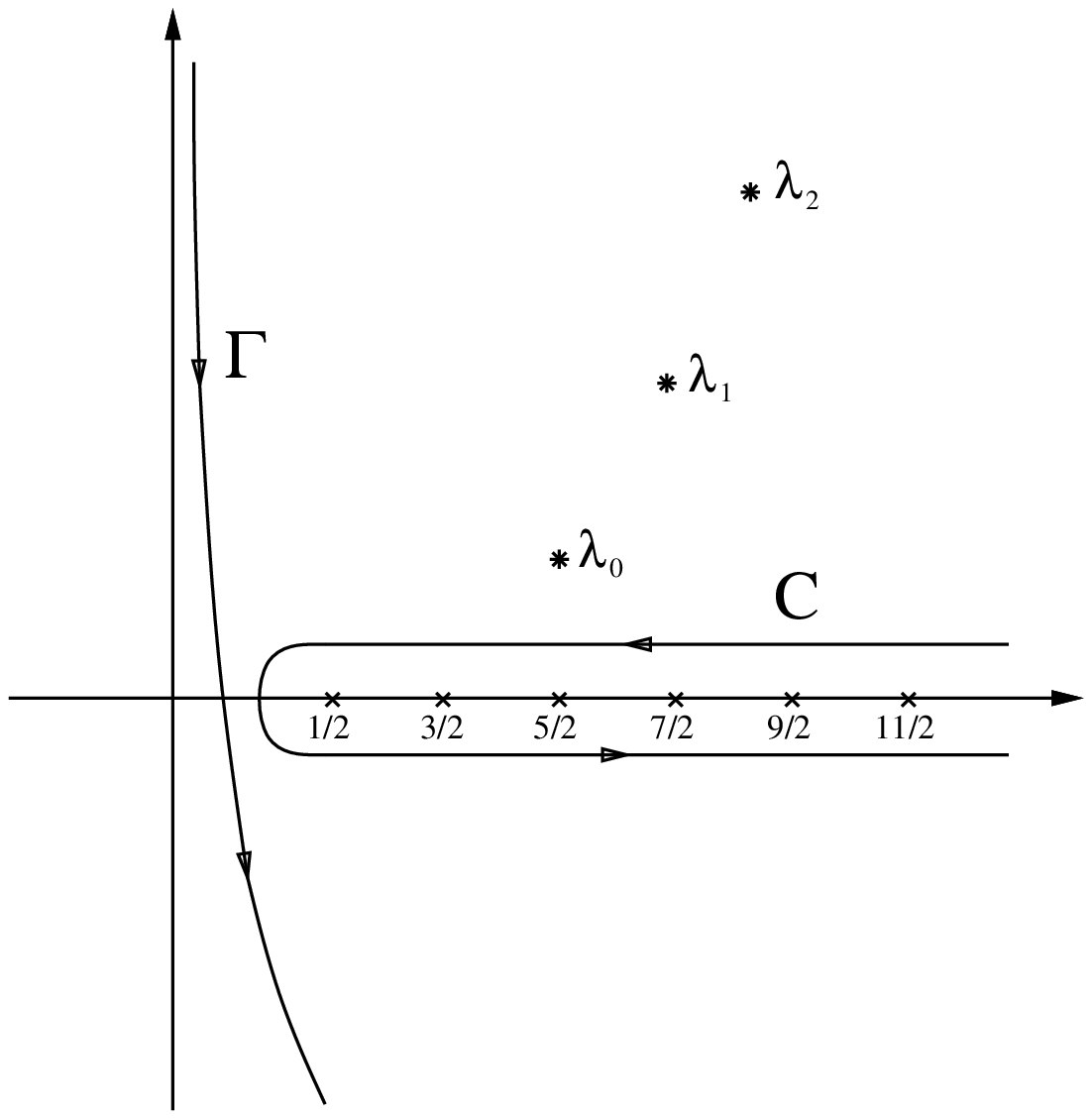}}
 \def\deflection{
\epsfysize=6 cm
 \epsfbox{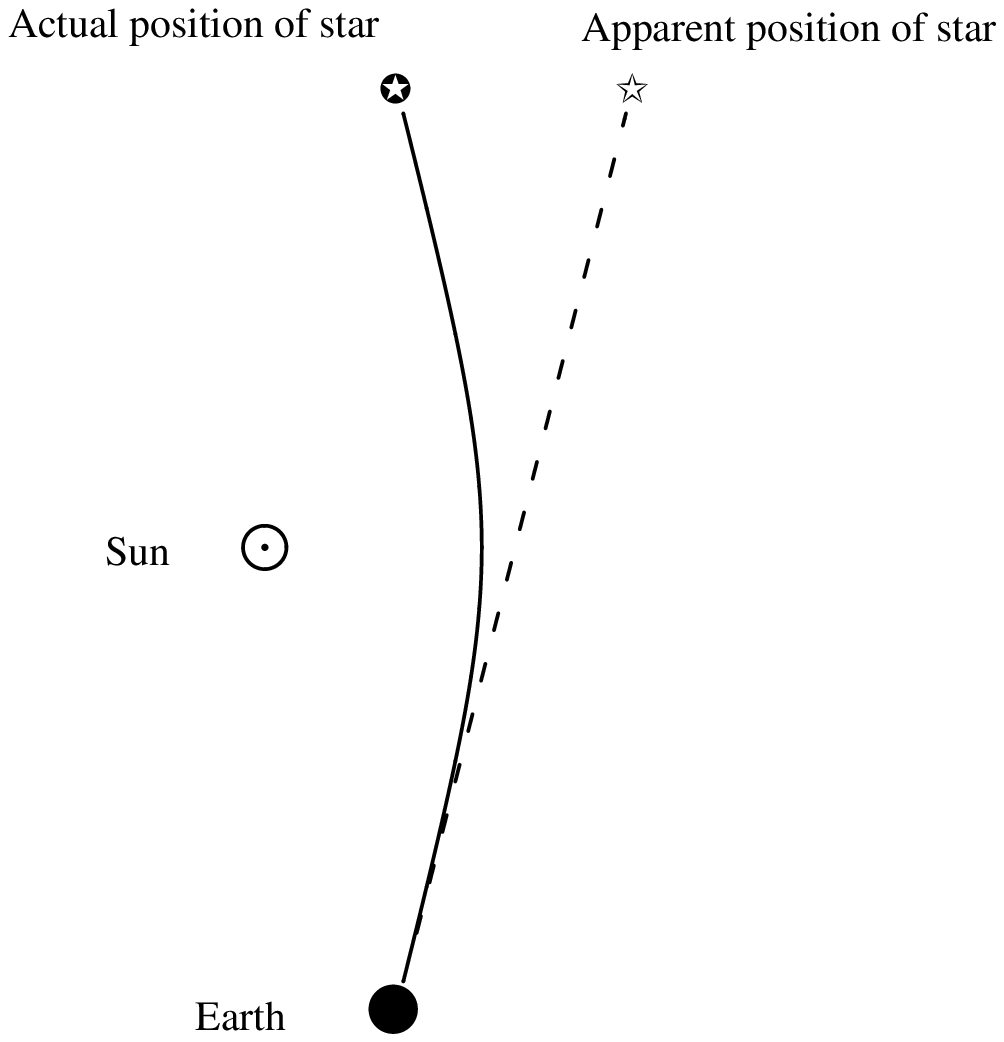}}

\documentclass[eqsecnum,aps,prd,amsfonts,amssymb,twocolumn]{revtex4}

\usepackage{epsfig}

\begin{document}

\title{Scattering by Black Holes}

\author{Nils Andersson}

\address{Department of Mathematics\\ University of Southampton, %
 Southampton SO17 1BJ,
 UK}

\author{Bruce Jensen}

 \address{Department of Mathematics, University of Southampton,\\and  Marconi %
 Communications,\\ 551-553 Wallisdown Road,\\ Poole BH12 5AG, UK}




\begin{abstract}
This article is a slightly modified version of the authors' contribution
to {\em Scattering}, edited by Roy Pike and Pierre Sabatier, to be published by
Academic Press.
\end{abstract}

\maketitle

\section{What is a Black Hole?}

Black holes, objects so compact that not even light can escape
their gravitational pull, are among the most intriguing concepts
of  modern science. As Kip Thorne wrote in 1974:
``Of all the conceptions of the human mind from unicorns to gargoyles
to the hydrogen bomb perhaps the most fantastic is the black hole: a
hole in space with a definite edge over which anything can fall and
nothing can escape; a hole that curves space and warps time.''
Evidence that these exotic objects exist in
our universe has been mounting since the discovery 
of the archetypal black hole Cygnus X1 in 1971.
Today, the presence of supramassive black holes
(several million times as massive as our sun)
at the centre of many galaxies, and smaller black holes 
(5-10 times as massive as the sun) in X-ray binary systems,
is generally accepted.  

From the mathematical point of view, a black hole is a spacetime defined 
by a four-dimensional metric tensor, that is, a solution
of Einstein's equations, with characteristic properties.
The simplest black hole is spherically symmetric and
non-rotating. It is known as the Schwarzschild black hole,
and is described by the metric tensor  $g_{ab}$ which is given 
by 
\begin{eqnarray}
ds^2  &=& \sum_{a=1}^4 \sum_{b=1}^4 g_{ab}dx^a \, dx^b \equiv  
g_{ab}dx^a \, dx^b = \nonumber \\
  &=& - {\Delta \over r^2} dt^2 + {r^2 \over \Delta} dr^2  + 
r^2 d\theta^2 + r^2 \sin^2 \theta \, d\varphi^2
\label{schw}
\end{eqnarray}
\index{black hole!spherical}
Here $ds$ is the infinitesimal element of proper time,
the coordinates are $\{x^a\}= (t,r,\theta,\varphi)$,
and we have defined $\Delta = r^2-2Mr$ where
$M$  is a constant given by $M= G M_{BH} /c^2$ where $M_{BH}$ is the black hole
mass and $G$ is Newton's constant. In the following we will always choose
so-called geometrised units in which both  the
speed of light $c$ and $G$ are equal to unity. We will also
assume that repated indices indicate summation, as in (\ref{schw}).

We are interested in scattering problems involving black holes. 
In such problems the curvature of spacetime
enters not only at the level of the boundary conditions, but 
also in the equations describing the propagation of the
various wave-fields (scalar, electromagnetic or gravitational)
that we may be interested in. 
Therefore the problem 
of scattering by black holes has more in common with scattering
in media with a non-constant index of refraction than scattering
by a physical object.
In the case of a black hole,
it is the curvature of space-time itself which is doing the scattering.

In this article we aim to summarise work in this research area,
and relate them to results in  familiar contexts such as quantum
scattering. We will discuss  possible diffraction effects, 
resonances, and digress on some of the peculiarities
of the black-hole problem. It is useful to begin our
discussion with the simplest of scattering problems involving black holes;
namely, the bending of light rays propagating
in a black hole spacetime. 

\section{Classical  Trajectories}
\subsection{Photon trajectories outside Schwarzschild black holes}

\begin{figure}[h]
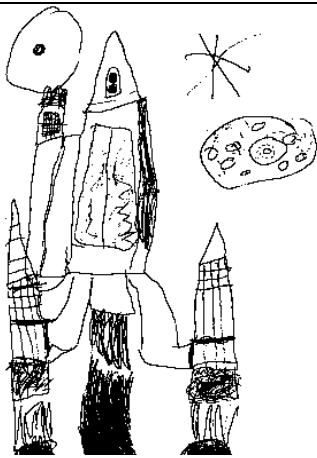

\caption{An astronaut maneuvers his rocket ship near
a non-rotating black hole.}
\hrule
\rocket
\smallskip
\hrule
\label{rocket}
\end{figure}

Let us consider an
astronaut who, while piloting his spaceship towards 
a Schwarzschild black hole (Figure \ref{rocket}),
shines a laser directly out his window, in 
the positive $\varphi$ direction.
The trajectory of the laser beam can be found by solving the equation for
a null geodesic (a line that is ``as straight as possible'' in the curved spacetime). Hence we are looking for a solution to the geodesic equation 
\begin{equation}
{d \over d\lambda} \left( {dx^\alpha \over d\lambda} \right) 
+ \Gamma^\alpha_{\mu\nu} {dx^\mu \over d\lambda} {dx^\nu \over d\lambda}
=0
\end{equation}
where the Christoffel symbols $\Gamma^\alpha_{\mu\nu}$ are functions of
the coordinates $\{ x^\alpha\}$. In order to be null, our geodesic
must satisfy $x^\mu x_\mu = 0$. 
Now, if we use the Schwarzschild coordinates introduced in 
(\ref{schw}) we can cast the equation for a null 
geodesic into the following form:
\begin{equation}
{d\varphi \over dr} = \pm {1\over r^2} \left[ {1 \over b^2} -  {1\over
r^2}\left( 1 - {2M\over r}\right) \right]^{-1/2} \ .
\label{nulgeoeq}
\end{equation}  
where $b$ is the impact parameter defined by $b=L/E$ where 
$L$ and $E$ are the angular momentum and the energy associated with 
the photon. From equation (\ref{nulgeoeq}) we can  deduce 
the properties of various light trajectories. 

\begin{figure}[h]
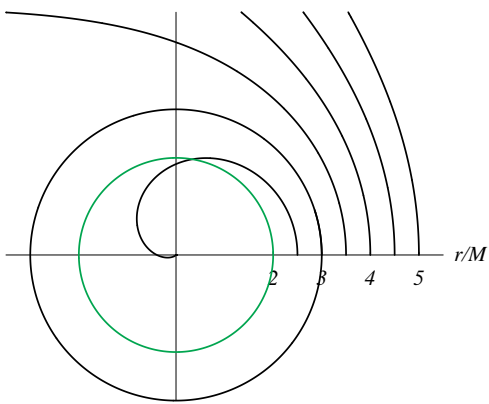

\caption{Light trajectories in the Schwarzschild geometry for various
values of the impact parameter $b$. 
This shows what happens to a laser beam which shines out the window
of a spaceship as it plunges into a (Schwarzschild) black hole.
Note the circular `photon orbit' at $r=3M$. The (grey) circle
at $r=2M$ represents the event horizon of the black hole.}
\hrule
\nulgeos
\smallskip
\hrule
\label{nulgeos}
\end{figure}

As the astronaut nears the black hole, the laser
beam is deflected more and more by the spacetime 
curvature --- see Figure~\ref{nulgeos}. 
The curves in Figure~\ref{nulgeos} are the solutions of equation 
(\ref{nulgeoeq})
with $\varphi'(b)=0$ and $b$ ranging from  $2.5M$ to $5M$.
It should be noted that $r=3M$ corresponds to a circle. This is
known as the unstable photon orbit, and its existence means that 
our astronauts laser beam will circle the black hole and illuminate
his neck! 
After passing  $r=3M$ the
astronaut 
finds that the laser beam is always deflected into the black hole.
 
\begin{figure}[h]
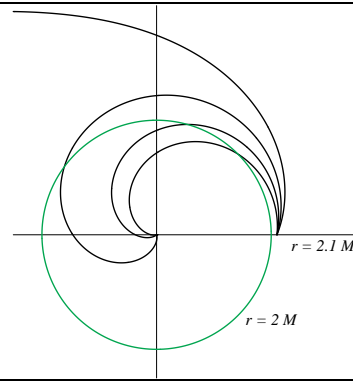

\caption{More light trajectories in the Schwarzschild spacetime. 
At $r=2.1 M$, the astronaut finds that a light beam from his 
spaceship can escape the black hole for only a small range of
angles. As $r\to 2M$, this becomes a single point.}
\hrule
\anglesp
\hrule
\label{anglesp}
\end{figure}

When he reaches $r=2.1 M$, the astronaut applies his rockets in such a way 
that the spaceship hovers at a constant distance from the black hole, 
and tries to
shine his light at various angles. He then
finds  that there is still a (small)  range of
angles at which the light beam can escape the black hole 
(see Figure~\ref{anglesp}). 

Once the astronaut has resumed his fall towards the black hole, 
and reached $r\to 2M$, the solid angle into which he 
must shine his laser in order for the light to escape the black 
hole has shrunk to a single point. He must aim it in 
the positive $r$ direction.  However, even if he hovers at constant $r=2M$, 
the ratio of the light frequency received
($\nu^\prime$) by his home planet at (say) $r'=\infty$ to the frequency 
emitted ($\nu$) is
 the ratio of the proper times at the two points. This follows from
the formula for the {\em gravitational redshift}:
\begin{equation}
\nu' = \nu (1-2M/r).
\end{equation}
From this we see that when $r\to 2M$ the light is completely redshifted away. 
Therefore any light emitted after
the astronaut reaches this limit, whatever the direction,  
remains inside the black hole. Our space traveller has reached the so-called
event horizon, and he can neither escape the black hole nor 
alert a rescue team of his fate.

\subsection{Bending of Starlight}

The above discussion illustrates some of the extreme effects that
the curvature of a black-hole spacetime might have on light trajectories.
Still, the ideas are relevant also in a  more familiar setting. 
In fact, the first experimental verification of Einstein's
theory of 
general relativity was the measurement of
the bending of starlight by the gravitational
field of the sun during a solar eclipse of 1919. 
While the sun is certainly not a black hole, the metric
tensor exterior to the surface of the sun is accurately described by 
(\ref{schw}). Thus, in a sense, the first
test of general relativity was 
also the first `black-hole scattering' 
experiment.

\begin{figure}[th]
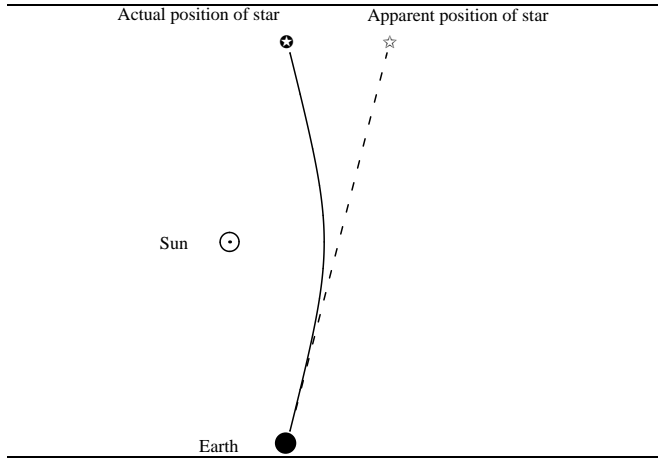

\caption{The earliest `black-hole scattering'
experiment: deflection of starlight by the sun.}
\hrule
\deflection
\hrule
\label{deflection}
\end{figure}

Let us assume that $M/r$ is small
($<<1$) 
but not negligible. If we introduce a new variable
\begin{equation}
y =   {1\over
r}\left( 1 - {M\over r}\right) \ ,
\end{equation}
we can  approximate our equation (\ref{nulgeoeq}) as
\begin{equation}
{d\varphi \over dy} = { 1+2My \over \left( {1 \over b^2} -y^2
\right)^{1/2} } + O(M^2/r^2) \ . 
\end{equation}  
Integration of this leads to
\begin{equation}
\varphi - \varphi_0 = {2M \over b} + \sin^{-1} (by) - 2M \left( 
{1 \over b^2} -y^2 \right)^{1/2} \ . 
\end{equation}
The initial trajectory was  such that $y\to 0$, which means
that
$\varphi \to \varphi_0$, defining the incoming direction. 
The smallest value of $y$ that the
light ray will ever reach corresponds to $y = 1/b$ when
$\varphi - \varphi_0 = 2M/b + \pi/2$. By symmetry  the
ray
will be deflected by an equal amount as it recedes to infinity. 
Hence, the final result will be   $\varphi - \varphi_0 = 4M/b + \pi$, 
and since the result would have been   
 $\varphi - \varphi_0 =  \pi$ if the light ray had followed a straight
path, we have  a total deflection of
\begin{equation}
\Delta \varphi = {4M \over b} \ .
\end{equation}
This result was first calculated by Einstein, and 
as noted above, this deflection was measured by Eddington and
collaborators in 1919.

\def\sqr#1#2{{\vcenter{\vbox{\hrule height.#2pt
       \hbox{\vrule width.#2pt height#1pt \kern#1pt
          \vrule width.#2pt}
        \hrule height.#2pt}}}}
\def\frac#1/#2{\textstyle{#1\over #2}}

\def\square{{\mathchoice\sqr64\sqr64\sqr43\sqr33  \, }}
\def\smsquare{{\mathchoice\sqr53\sqr53\sqr42\sqr32  \, }} 

\subsection{Gravitational lensing}

\begin{figure}[th]
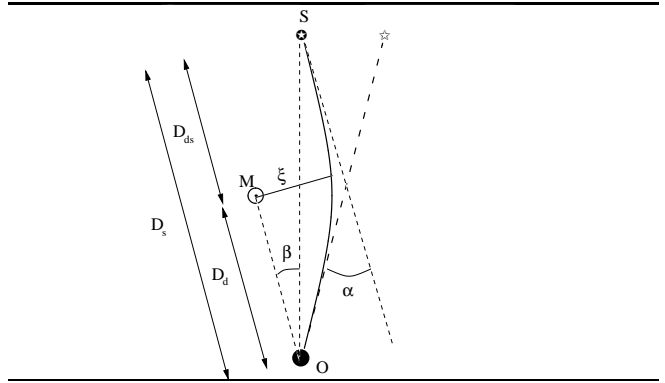

\caption{A schematic description of the simplest gravitational 
lensing geometry. Light rays are bent as the pass by a point
source located at point $M$ on their way from the 
distant source at $S$ to the observer $O$.}\strut
\hrule
\glens
\hrule
\label{glens}
\end{figure}

With enhanced observational capabilities, ranging from
large radio telescopes to the repaired Hubble space telescope, 
astronomers have found plenty of evidence that light
often bends as it travels through space. This is known 
as gravitational lensing.
Before discussing some of these observations
it is instructive to describe the simplest lensing 
situation; the case when light 
from a distant quasar is lensed by a localised mass distribution
(treated as a point source with mass $M$) on its way 
to the observatory. The relevant geometry is shown in Figure~\ref{glens}.
We denote the true angular separation from the lens to be $\beta$,
consider a light ray that passes the lens at a minimum distance
$\xi$ (essentially the impact parameter from the previous section) 
and is deflected an angle $\alpha= 4M/\xi$ by the spacetime geometry.
Assuming that $\xi>>2M$ we do not need to worry about diffraction effects, 
and can analyse the problem using simple geometry. Disregarding 
complicating factors one can show that the condition that the 
light ray reaches the observer leads to
\begin{equation} 
\beta = \theta - \alpha {D_{\rm ds} \over D_{\rm s}} =
\theta - {\alpha_0^2 \over \theta}  
\end{equation}
where we have defined
\[
\theta = {\xi \over D_{\rm d}} \ , \qquad \alpha_0^2 = 4M {D_{\rm ds} \over 
D_{\rm d} D_{\rm s}}
\]
and the distances $D_{\rm ds}$,  
$D_{\rm d}$ and  $D_{\rm s}$ are as shown in Figure~\ref{glens}.

In other words, we  solve
\begin{equation} 
\theta^2 - \beta \theta - \alpha_0^2 = 0 
\end{equation}
and get
\begin{equation}
\theta_{1,2} = {1\over 2} \left( \beta \pm \sqrt{ 4\alpha_0^2 + \beta^2} \right)
\end{equation}
That is, we always get two solutions of opposite sign. This means that
there will typically be one image on each side of the lens. 
A special case worthy of notice arises when the source, lens and observer
are all aligned. Then we have $\beta=0$ and it is easy to realise that there
is then no preferred plane for the light rays to travel in. Thus 
the whole ring of angular radius $|\theta| = \alpha_0$ is a solution to 
the simple lensing equation. This is commonly known as an `Einstein ring' 
and, as is easy to see, 
such images can only occur in lensing by axially symmetric mass
distributions.

The above description is obviously idealised in many ways, and in 
analysing observed gravitational lenses one must consider
much more complicated mass distributions, as well as use
detailed cosmological
models. Still, the above example illustrates the basic
principles, and 
we now turn to some actual observations of gravitational
lensing.

The first lensing candidate was observed using a radio
telescope at Jodrell Bank in 1979 and
is catalogued as 0957+561. It is a typical example of a
double image of a distant quasar. When the spectra of the two images
were studied it was found that they were remarkably similar, 
but redshifted to slightly different extent. It was 
concluded that the two images were extremely
unlikely to correspond to separate individual quasars.
 
\begin{figure}[th]
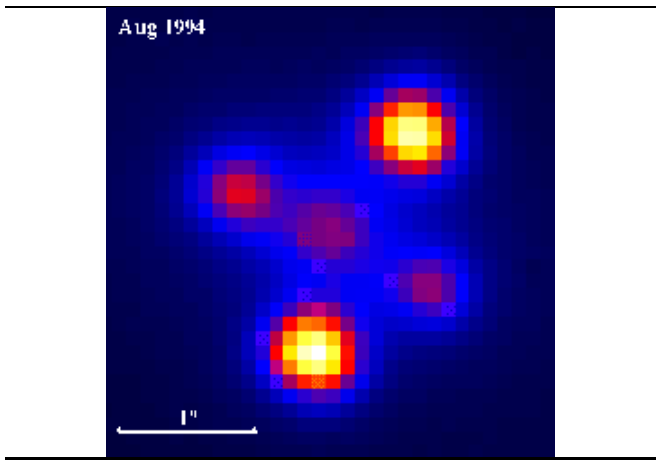

\caption{The famous "Einstein Cross" is a case where a gravitational 
lensing by a massive galaxy (central image)
leads to four distinct images of 
a distant quasar. This picture was taken by the William Hershel Telescope
in August 1994.
}\strut
\hrule
\einstein
\hrule
\label{einstein}
\end{figure}

Since this first discovery, many other double image 
systems have been observed.
More complicated systems, comprising of further images, have
also been found. Among them are the triple image lenses
2016+112 and 0023+171. Perhaps the most unique 
multiple image case 
found so far is 2237+0305, more commonly known as the 
``Einstein cross''. In this beautiful system, shown in 
figure~\ref{einstein},  one can see
four distinct images.

The above examples are all likely examples of lensing by 
`point sources'. For extended mass distributions, like 
galaxies, one would expect to see also arcs and in some
unique cases almost complete Einstein rings. The first arcs
(Abell 370 and Cl224-02) were actually found when 
gravitational lenses were still considered a mere
theoretical possibility. Since then many further examples
of lensed arcs have been discovered, as well as nearly
complete rings. One interesting example of the latter is
MG1131+0456. 

As our observational capabilities continue to improve the
list of lensed systems is rapidly growing. New observations
provide challenges for the theorists that want to 
deduce the geometry of the lensing mass distribution as well
as understand the nature of the 
original light source. Ideally, one would also like to be
able to use lensing observations of distant quasars to 
also deduce information about cosmology. Considerable 
progress in these directions have been made in recent years, 
and our understanding should continue to improve with the 
observational data.  For further details, we refer the 
reader to the monograph by Schneider, Ehlers and Falco \cite{schneider}.

\section{Wave scattering}

Having discussed some classic examples and exciting  
observations of the scattering of light by massive bodies
we now turn to the issue of possible diffraction effects. 
Essentially, most models of the gravitational lensing
phenomenon are based on ``geometrical optics''.
Therefore one would not expect these calculations to yield
any insight into possible wave phenomena. However,
as we will see, the 
extreme nature of black holes lead to the existence of many 
complicated diffraction effects. To understand these it is 
essential that we develop a framework for studying the 
scattering of waves by black holes. From an observational
point of view, our main interest will be focussed on 
electromagnetic and gravitational waves. The latter 
are particularly interesting since a new generation 
of gravitational-wave detectors is due to come online in 
the next few years. It is generally believed that these
will make the long-heralded field of gravitational-wave astronomy
a reality, and that they will allow us ot make detailed
observations of the physics in the immediate vicinity
of a black hole \cite{kip}. 

In order to introduce the various concepts involved in studies
of the scattering of waves by black holes we will consider the
relatively simple case of scalar waves. This may 
seem like a peculiar choice given that no massless scalar 
fields have yet been observed in nature. However, it turns
out that the main equations governing a weak electromagnetic field,
or gravitational waves, in a curved spacetime are essentially 
the same as the scalar field wave equation (see for example equation
(\ref{teuk0})). Hence, the scalar field serves
as a useful model.  See Chandrasekhar \cite{chandra} for an exhaustive study.

\subsection{Scalar Fields in the Schwarzschild geometry} 

Let us consider a scalar field $\Phi$ propagating in the 
Schwarzschild spacetime, as described by (\ref{schw}). 
The equation governing the evolution of the scalar field is 
\begin{equation}
\square \Phi = {1\over \sqrt{g} }
{\partial \over \partial x^\mu} \sqrt{g}
g^{\mu\nu}
{\partial \over \partial x^\nu} 
\Phi
=0
\end{equation}
where $g$ is the determinant of $g_{\mu\nu}$. Since the 
Schwarzschild spacetime is spherically symmetric, 
we may assume a Fourier decomposition
\begin{equation}
\Phi (x^\mu) = {1 \over 4\pi r} \sum_{l=0}^\infty \sum_{m=-l}^l e^{-i\omega t} 
Y_l^m(\theta,\varphi) \hat{\phi}_{lm}(\omega,r)
\end{equation}
where $Y_l^m$ are the spherical harmonics. If the boundary
conditions are cylindrically symmetric, as they would be in the
plane-wave scattering problem, the $\varphi=0$ axis ($z$-axis)
can always be chosen to be the axis of symmetry. We can therefore
assume that $\hat{\phi}_{lm}$ to be independent of $\varphi$ and 
perform the sum over $m$, writing
\begin{equation}
\Phi (x^\mu) =  \sum_{l=0}^\infty (2l+1)  e^{-i\omega t} 
P_l(\cos \theta) {\hat{\phi}_{l}(\omega,r) \over r}
\label{sdec}
\end{equation}
where $P_l$ is a Legendre function.

\begin{figure}[h]
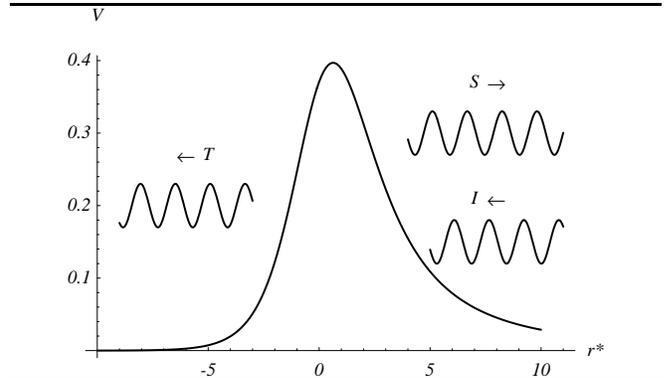

\caption{A schematic description
of the scattering of waves in the Schwarzschild
background. The effective potential of equation
(\ref{de}) is shown as a function of $r_*$. The event 
horizon of the black hole is located at $r_\ast=-\infty$. 
An incident wave $I$ is decomposed into a transmitted
component $T$ and a scattered component $S$. 
}\strut
\hrule
\hump2
\hrule
\label{hump2}
\end{figure}

In the Schwarzschild spacetime the wave equation for the 
scalar field reduces 
to the following Schr{\"o}dinger-type equation for
$\hat{\phi}_l$: 
\begin{equation}
{d^2 \hat{\phi}_l \over dr_*^2} + 
\left[ \omega^2 - V(r)\right] \hat{\phi}_l =0,
\label{de}
\end{equation}
where the so-called tortoise coordinate is defined by
\begin{equation}
{d \over dr_* } = {r-2M \over r} {d\over dr}
\end{equation}
This integrates to
\begin{equation}
r_* = r + 2M \log\left( {r\over 2M}-1 \right) 
\end{equation}
and we see that introducing the tortoise coordinate 
corresponds to ``pushing  the event horizon of the black hole
away to $-\infty$''. The effective potential is explicitly
given by
\begin{equation}
V(r) = {r-2M \over r} \left[ {l(l+1) \over r^2} + {2M\over r^3}\right].
\end{equation}
It is positive definite 
and has a single peak in the range $r_\ast\in [-\infty,\infty]$, see
Figure~\ref{hump2}.

\begin{figure}[h]
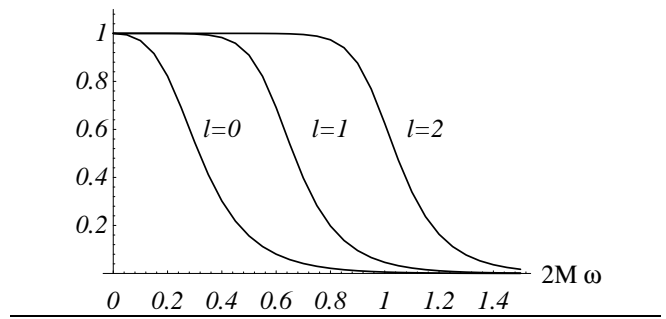

\caption{Scalar wave scattering coefficient $S_l$ for $l=0,1,2,3$ as
a function of $2M \omega$.}
 \hrule
 \scat
 \hrule
 \label{scat}
 \end{figure}

A black hole is distinguished by the 
fact that no information can escape from within the event horizon. 
Hence, any physical solution to (\ref{de}) 
must be purely ingoing at the event horizon, that is,
at $r=2M$ ($r_* \to -\infty$). 
Therefore we seek solutions to (\ref{de}) of form (for a 
given frequency $\omega$)
\begin{equation}
\hat{\phi}_l \sim \left\{ \begin{array}{ll} 
e^{-i\omega r_* + il\pi/2} - S_l(\omega) e^{+i\omega r_* -
il\pi/2} \qquad &  r_* \to \infty \\
T_l(\omega)e^{-i\omega r_* }
\qquad \qquad \qquad\qquad & r_* \to -\infty
\end{array} \right.
\label{bc1}
\end{equation}
where the amplitudes of the scattered and the transmitted waves,  
$S_l$ and $T_l$, remain to be determined.  
Clearly, problems involving waves scattered from a Schwarzschild
black hole share many features with scattering problems in quantum 
theory. Hence, we can adopt standard techniques to evaluate $S_l$ and
$T_l$. By conservation of flux it follows that
\begin{equation}
|T_l|^2 = 1 - |S_l|^2
\end{equation}
Hence, we need only determine either $S_l$ or $T_l$. 
Typical results for $S_l$ are shown in Figure~\ref{scat}. 

The nature of $S_l$  can be understood from the following observations. 
For $\omega<<2M$ the wavelength of the infalling wave is so large that 
the wave 
is essentially unaffected by the presence of the black hole. It is only
if we ``aim'' the wave straight at the black hole (recall that the 
impact parameter follows from $b=L/E\sim l/\omega$)
that we can get an appreciable effect. Hence, we expect to 
have $S_l \to 1$ as $\omega \to 0$. For large frequencies
$\omega>>2M$, the situation is the opposite and we expect to 
find that $S_l\to0$ as $\omega \to \infty$. Thus, high 
frequency waves will  be absorbed unless they are aimed away
from the black hole.

\begin{figure}[h]
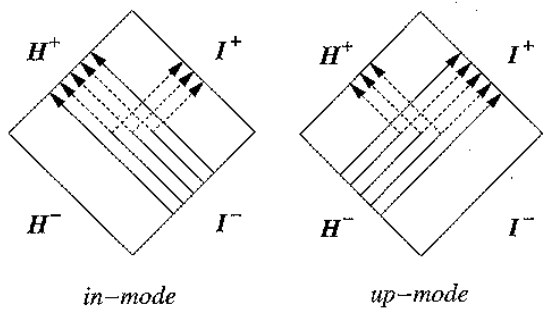

\caption{The nature of two linearly independent
solutions to the scalar wave equation outside a black hole.
The in-mode corresponds to purely ingoing waves crossing the
 event horizon ($H^+$), while the up-mode 
corresponds to purely outgoing waves at  spatial infinity
($I^+$).
}
\hrule
\inup
\hrule
\label{inup}
\end{figure}

In studies of dynamical black holes one often needs to construct the general 
solution to (\ref{de}). One can do this using two linearly
independent solutions. These are customarily normalised in a 
way that differs slightly from (\ref{bc1}). A first solution
(essentially (\ref{bc1})) is such that the amplitude of the 
waves falling across
the event horizon are normalised to unity, and one requires the
amplitudes of out- and ingoing waves at infinity. 
This solution is sometimes called the ``in''-mode and it can be written 
\begin{equation}
\hat{\phi}_l^{\rm in} \sim \left\{ \begin{array}{ll}
e^{ -i\omega  r_\ast}\ , & r_\ast\rightarrow  -\infty \ , \\
A_{\rm out}(\omega)e^{ i\omega  r_\ast}+A_{\rm in} (\omega) 
e^{ -i\omega  r_\ast}\ , & r_\ast \rightarrow   
+\infty \ ,
\end{array}
\right.
\label{inmode}\end{equation}
Given this solution, a second linearly independent one
corresponds to waves of unit amplitude reaching spatial infinity.
This is the ``up'' mode, and it follows from 
\begin{equation}
\hat{\phi}_l^{\rm up} \sim \left\{ \begin{array}{ll}
B_{\rm out}(\omega)e^{ i\omega  r_\ast}+B_{\rm in}(\omega)e^{ -i\omega  
r_\ast}\ ,
&r_\ast \rightarrow   
-\infty \ ,  \\
e^{ +i\omega  r_\ast}\ , & r_\ast\rightarrow  +\infty 
\end{array}
\right. ,
\label{upmode}\end{equation}
 The nature of these two solutions is illustrated in Figure~\ref{inup}.

For those unfamiliar with conformal diagrams,
a short explanation of Figure~\ref{inup} is necessary. 
Suppressing the angular coordinates $(\theta, \phi)$, we 
make a coordinate transformation in $(t, r)$ such that the half-infinite
space exterior to the black hole is mapped onto a finite portion of the plane.
The transformation is chosen so that the light cones always intersect
at an angle of 45$^\circ$ ---much like Mercator's projection of the earth distorts the
shape of the continents but preserves the directions North-South and
East-West. The four `points at infinity' are mapped into the diagonal edges as
follows:
\begin{table}[h]
\begin{tabular}[t]{lll}
Name & Symbol &  Coordinates\\
\hline
Past Horizon & $\hbox{\bf H}^-$ &$r^* = -\infty$, $t=-\infty$ \\
Future Horizon & $\hbox{\bf H}^+$ & $r^* = -\infty$, $t=+\infty$ \\
Past Null Infinity & $\hbox{\bf I}^-$ & $r^* = +\infty$, $t=-\infty$ \\
Future  Null Infinity & $\hbox{\bf I}^+$ & $r^* = +\infty$, $t=+\infty$ \\
\end{tabular}
\label{scris}
\end{table}
The diamond-shaped figures
represent the whole of the space-time exterior to the black hole. The
various arrows
represent the path followed by the incident, transmitted 
and reflected wave fronts.  

\subsection{Plane wave scattering}

In order to better understand the physics of black holes 
we  want to formulate a scattering problem
analogous to that used to probe the nature of (say) 
nuclear particles. We want to let a plane wave fall
onto the black hole, and investigate how the black hole
manifests itself in the scattered wave.
This is obviously a model problem, given that we
cannot expect to ever be able to compare the calculated scattering
cross sections to real observations, but it is still  
instructive. In particular, it will lead to an
unveiling of a deep analogy between black hole physics
and well known phenomena such as glory scattering.  

However, in formulating this problem we immediately face
difficulties.   What exactly do we mean by a {\em plane wave} 
in a curved spacetime? It turns out that we can answer this 
question by appealing to the  analogous problem of 
Coulomb scattering. As with the charge in the Coulomb problem, the black
hole contributes a long-range potential that falls off as $1/r$ 
at large distances. The effect of such a 
long-range potential on the ``plane'' wave 
can be accounted for by a simple modification of the
standard (flat space) expressions for the 
scattering amplitude. In the black-hole case we essentially need
to introduce $r_\ast$ in the phase of the plane wave, and we obtain
\begin{equation}
\Phi_{\rm plane} \sim {1 \over \omega r} \sum_{l=0}^\infty 
i^l (2l +1) P_l (\cos \theta) \sin \left[ \omega 
r_\ast - {l \pi \over 2} \right] 
\label{plane}\end{equation}
as $r_\ast \to +\infty$ in the case of scalar waves.

\subsection{Phase-shifts and deflection function}

Having defined a suitable plane wave, 
the scattering problem involves finding a
solution to (\ref{de}) , {\em i.e.}, identifying 
the asymptotic amplitudes $A_{\rm in}$ and
$A_{\rm out}$ for a given frequency $\omega$. 
Then we can extract the scattered wave by discarding the
part of the solution that corresponds to the original plane wave. The
physical information we are interested in is contained within the
scattering amplitude $f(\theta)$, which follows from
\begin{equation}
\Phi \sim \Phi_{\rm plane} + {f(\theta) \over r} e^{i\omega r_\ast} ,
 \quad {\rm as}\  r_\ast \to +\infty \ .
\end{equation}

Now letting
\begin{eqnarray}
\Phi  - \Phi_{\rm plane} \sim {1\over 2i\omega r} e^{i\omega r_\ast} &&\sum_{l=0}^\infty (2l
+1)\left[ e^{2i\delta_l} - 1 \right]  
P_l (\cos \theta)  \nonumber \\
&& \quad {\rm as}\  r_\ast \to +\infty \ ,
\label{phidiff}\end{eqnarray}
define the (complex-valued) phase-shifts $\delta_l$ it 
is straightforward to show that
\begin{equation}
 e^{2i\delta_l} = S_l = (-1)^{l +1} {A_{\rm out} \over A_{\rm in}} \ .
\label{psh}\end{equation}
From this it follows that the scattering amplitude, 
that contains all the physical information, is given by
\begin{equation}
f(\theta) = {1\over 2i\omega} \sum_{l=0}^\infty (2l +1)
\left[ e^{2i\delta_l} - 1 \right]  P_l (\cos \theta) \ .
\label{scamp}\end{equation}

When discussing the physical quantities that follow from a set of
phase-shifts it is natural to use Ford and Wheeler's excellent
description of semiclassical scattering  from 1959~\cite{ford}.
In the semiclassical picture the
impact parameter $b$ is given by
\begin{equation}
b = \left( l + {1 \over 2} \right) {1 \over \omega} \ .
\label{impact}\end{equation}
Then each partial wave is considered as impinging on the 
black hole from an initial distance $b$ away from the axis.

\begin{figure}[h]
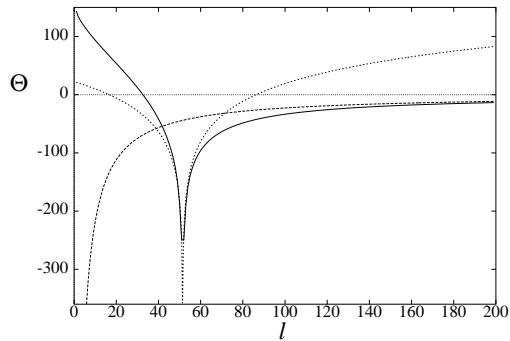

\caption{ The deflection function  $\Theta$ (solid) is shown as a function of
$l$ for $\omega M =10$. For large impact parameters (large $l$)
the approximate results approach the Einstein deflection angle $-4M/b$
(dashed). A logarithmic singularity in $\Theta$ is apparent at the
critical impact parameter ($l_c \approx 51.5$ here). This feature is
associated with the unstable photon orbit at $r=3M$. Also shown (as a dashed curve) is an approximation obtained by inverting Darwin's formula.
}
\hrule
\deflect
\hrule
\label{deflect}
\end{figure}

In this description much physical
information can be extracted from the so-called deflection function
$\Theta (l)$. 
 It corresponds to the angle by which a certain partial wave is
 scattered by the black hole, and is
related to the real part of the phase-shifts;
\begin{equation}
\Theta (l)= 2 {d \over dl} {\rm Re}\ \delta_l \ .
\end{equation}
Here $l$ is allowed to assume continuous real values.  
We can obtain approximations for $\Theta(l)$ in
some limiting cases. For large values
of the impact parameter, $b$, one would expect the value of the
deflection function to agree with Einstein's classic result $\Theta \approx
- 4M/b$, that we derived earlier. 
As can be seen in Figure~\ref{deflect} this is certainly the
case. A numerically determined $\Theta(l)$ rapidly approaches 
the approximate result as $l$ increases.

A second approximation
is intimately related to the existence of a glory in black-hole
scattering.  
Whenever the classical cross section diverges in either the forward or
the backward direction a diffraction phenomenon called a glory arises.
This phenomenon is well-known in both optics and
quantum scattering.  In general, backward glories can occur if
$\Theta < - \pi$ for some values of $b$. Whenever the deflection
function passes through zero or a multiple of $\pi$ we have a glory. In
the black-hole case one would expect glory scattering to be associated
with the unstable photon orbit at $r=3M$. This
essentially means that we would expect a logarithmic singularity in
the deflection function to be associated with the
critical impact parameter $b_c = 3 \sqrt{3} M$. This feature is obvious
in Figure~\ref{deflect}. Many years ago Darwin~\cite{darwin} deduced an
approximate relation between the impact parameter and the deflection
function close to this singularity;
\begin{equation}
b(\Theta)  \approx 3\sqrt{3}M  + 3.48Me^{-\Theta} \ .
\label{darwin}\end{equation}
If we invert this formula and use (\ref{impact}) we get $\Theta$ as
 a function of $l$. As can be seen in Figure~\ref{deflect} this 
approximation
 is in excellent agreement with the deflection function obtained from
 the approximate phase-shifts.

\subsection{The black-hole glory}

\begin{figure}[h]
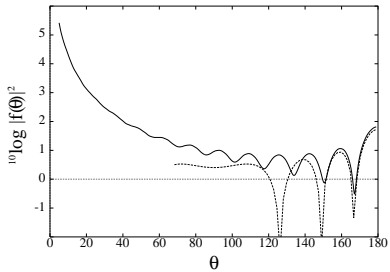

 \hrule
\cross
 \hrule
\caption
{ The scalar wave differential cross section for $\omega M = 2.0$.
The oscillations on the backward direction corresponds to the 
black-hole glory. Also shown is an approximation to the glory
oscillations following from equation (\ref{glapp}).}
\label{cross}
\end{figure}

We now want to proceed to calculate the scattering amplitude through
the partial-wave sum  (\ref{scamp}). In doing this, we must 
proceed with caution since, as in the Coulomb problem, 
the sum is divergent. This problem is normally
 avoided  by introducing a cutoff where the 
remainder of the true
partial-wave sum is replaced by analytic results for a limiting case.
 In essence, we extract the
contribution from large impact parameters from (\ref{scamp}) , {\em i.e.},
replace it by
\begin{equation}
f(\theta) = f_N(\theta) + f_D(\theta) \ ,
\label{scpract}\end{equation}
where the long-range (Newtonian) contribution is given by
\begin{equation}
f_N(\theta) = M {\Gamma(1-2iM\omega) \over \Gamma(1+2iM\omega)} 
\left[ \sin {\theta \over 2} \right]^{-2+4iM\omega} \ ,
\label{scexact}\end{equation}
and
\begin{equation}
f_D(\theta) =  {1\over 2i\omega} \sum_{l=0}^\infty 
(2l +1)\left[ e^{2i\delta_l} - e^{2i\delta_l^N} 
\right]  P_l (\cos \theta) \ ,
\label{2ndterm}\end{equation}
is the part of the scattering amplitude that  gives rise
to diffraction effects. The Newtonian phase-shifts $\delta_l^N$ 
follow from (cf. the standard Coulomb expression)
\begin{equation}
e^{2i\delta_l^N} =  {\Gamma(l + 1 -2iM\omega) \over 
\Gamma(l+ 1+2iM\omega)} \ .
\end{equation}
The sum in (\ref{2ndterm}) is convergent, and we can readily
determine the desired physical quantities. 
 The  quantity of main physical
interest is the differential cross section --
the ``intensity''of waves  scattered into a certain solid angle. It
follows from the well-known relation
\begin{equation}
{d\sigma \over d\Omega} = \vert f(\theta) \vert^2 \ .
\end{equation}
A typical example of a 
scalar wave cross section is shown in Figure~\ref{cross}. 
This provides a beautiful example of the black-hole glory (the
regular oscillations at large angles).
  
It is appropriate to discuss the details of the cross section
shown in Figure~\ref{cross} in somewhat more detail. 
First of all, one would expect the long-range attraction of the
gravitational interaction to give rise to a divergent focusing ($\sim
\theta^{-4}$) in the forward direction.  Secondly, 
we expect to find that interference between
partial waves associated with the unstable photon orbit, {\em i.e.},
with impact parameters $b\approx b_c =  3\sqrt{3}M$, gives rise to a
glory effect in the backward direction (cf. Figure~\ref{deflect}).
This effect is, of course, prominent in Figure~\ref{cross}. 
The glory oscillations in the cross section can be approximated by\cite{cdewitt}
\begin{equation}
\left. {d\sigma \over d\Omega} \right|_{glory} = 2\pi \omega b^2
 \left| {db \over d\theta} \right| J_0^2 (\omega b \sin \theta ) \ .
\label{glapp}\end{equation}
When combined with the Darwin formula (\ref{darwin}) this provides
a good approximation whenever $\omega M >> 1$ and $\vert \theta
- \pi \vert << 1$, see Figure~\ref{cross}.

According to the predictions of geometrical optics  one
might expect to find  glory oscillations not only in the 
backward direction, but
also in the forward direction. Partial waves associated with the
critical impact parameter may be deflected any multiple of $\pi$ and so
give rise to diffraction close to both $\theta =0$ and $\pi$. Moreover,
in Figure~\ref{deflect} we see that the deflection function  passes
through $\Theta = 0$ for a value of $l$ lower than that associated
with the unstable photon orbit ($l_c$). This means that there should be also a
forward glory. However, this feature drowns in the
divergence of the cross section that is due to the large $l$ partial
waves.  Moreover, partial waves corresponding to $l < l_c$ are to a large extent absorbed by the black hole. Hence, the
forward glory is  exceptionally faint.


\section{Time-dependent scattering --- Resonances}

As we have already pointed out, 
the scattering theory of monochromatic plane waves by a black hole, 
though interesting,
is not likely to have any experimental confirmation soon. 
The only possibility (as we can see it)
would be via diffraction effects observed in 
gravitational lenses. In order to discuss potentially 
observable scattering effects
we must turn to  the closely related problem of the response of the 
black hole to an initial perturbation. 
The perturbation response, the black hole's ``fingerprints'', 
will only be observable via gravitational waves.
Such observations are still outstanding, but the should 
become reality in the next few years when large 
scale interferometric gravitational wave detectors
(LIGO, VIRGO, GEO600, TAMA300) come online. A typical
astrophysical scenario that would produce copious
amounts of gravitational waves is 
a black hole in the act of swallowing a neutron star at the endpoint
of binary evolution~\cite{kip}. 
 The perturbation of the black-hole gravitational field
would produce outgoing gravitational radiation which can be
detected far from the black hole. We shall see below that this
 radiation has a characteristic spectrum which indicates the
presence of the hole and can be used to deduce its mass and angular momentum. 
In other words, gravitational-wave observations will bring 
direct observations of the most elusive of our universe's inhabitants, 
the black holes. 

\subsection{A Green's function approach}

We are interested in modelling the response of a black hole to 
a prescribed initial perturbation. As a model problem 
we continue with the scalar field, and consider the evolution
of the field as an initial-value problem. 
That is, instead of prescribing the asymptotic character of the field (as in 
the plane-wave case) we suppose that
we are given a specified scalar field at
time  $t=0$ and  that we want to calculate the future
evolution of the field.  The 
time-evolution of a wave-field $\Phi_l(r_\ast,t)$
follows from
\begin{eqnarray}
\Phi_l(r_\ast,t) &=& \int G(r_\ast,y,t)\partial_t 
\Phi_l(y,0) dy \nonumber \\
&+&  \int \partial_t G(r_\ast,y,t) \Phi_l(y,0) dy \ ,
\label{Gevol}\end{eqnarray}
for $t>0$. 
The (retarded) Green's function is defined by
\begin{equation}
\left[ {\partial^2 \over \partial r_\ast^2}- 
{\partial^2 \over \partial t^2} - V(r) \right] 
G(r_\ast,y,t) = \delta(t)\delta(r_\ast-y) \ ,
\label{Geqn}\end{equation}
together with the condition $G(r_\ast,y,t) = 0$ for $t\le 0$
and the appropriate space boundary conditions. These
conditions follow from
\begin{eqnarray}
{\partial G \over \partial r_*} + i\omega G = 0 \ , \quad r\to 2M \\
{\partial G \over \partial r_*} - i\omega G = 0 \ , \quad r\to \infty \ .
\end{eqnarray}

\begin{figure}[ht]
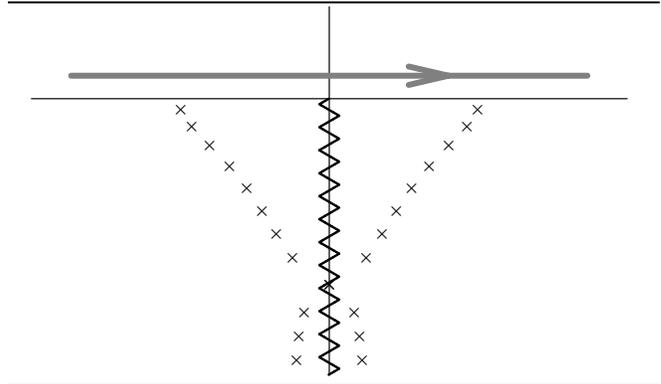

\caption{The integral contour of equation (\ref{greent}).
If $t>r_*$, the contour can be deformed and the integral
can be rewritten as a sum over the poles (crosses) and a remainder 
integral over the branch cut (zig-zag line).}
\hrule
\contour1
\smallskip
\hrule
\label{contour1}
\end{figure}

Let us take $\hat{G}$ to be the one-sided Fourier 
transform of  $G$:
\begin{equation}
\hat{G}(r_\ast,y,\omega) = \int_{0^-}^{+\infty}
G(r_\ast,y,t) e^{i\omega t} dt \ .
\label{ftrafo}\end{equation}
This transform is well defined as long as ${\rm Im}\ \omega \ge 0$,
and the corresponding inversion formula is
\begin{equation} 
G(r_\ast,y,t) = {1 \over 2\pi} \int_{-\infty+ic}^{+\infty+ic} 
\hat{G}(r_\ast,y,\omega) e^{-i\omega t} d\omega \ ,
\label{greent}\end{equation}
where $c$ is some positive number  (see Figure \ref{contour1}).
The Green's function $\hat{G}(r_\ast,y,\omega)$ can now be expressed in 
terms of
two linearly independent solutions to the homogeneous equation
(\ref{de}). The two required solutions are (\ref{inmode}) and 
(\ref{upmode}) as defined earlier, 
giving the Green's function:
\begin{equation}
\hat{G}(r_\ast,y,\omega) = -{ 1 \over 2i\omega A_{\rm in} (\omega)} 
\left\{ \begin{array}{lll} 
 \hat{\phi}_l^{\rm in}(r_\ast,\omega) \hat{\phi}_l^{\rm up}(y,\omega) 
 \ , & r_\ast < y \ , \\ \\
\hat{\phi}_l^{\rm in}(y,\omega) \hat{\phi}_l^{\rm up}(r_\ast,\omega) 
 \ , & r_\ast > y \ .
\label{greenf}\end{array} \right.
\end{equation}
Here we have used the Wronskian relation 
\begin{equation}
W(\omega) \equiv \hat{\phi}_l^{\rm in} {d \hat{\phi}_l^{\rm up} \over dr_\ast}
- \hat{\phi}_l^{\rm up} {d \hat{\phi}_l^{\rm in} \over dr_\ast}
= 2i\omega A_{\rm in} (\omega) \ . 
\label{wronsk}\end{equation}

\subsection{Quasinormal modes}

The initial-value problem can now, in principle, be 
approached by direct numerical
integration of (\ref{de}) for (almost) real values of $\omega$ and
subsequent inversion of (\ref{greent}). It has proved useful, however,
to deform the contour of integration in the complex $\omega$ plane
using Cauchy's theorem and rewrite the integral as a sum over residues
plus a remainder integral. Our first task in this process is to find the 
position of the poles of the Green's function.

The poles in the radial Green's function (\ref{greenf}) are located at
the simple zeros of $A_{\rm in}$, which we 
will denote $\{\omega_q\}$. When $\omega=\omega_q$, the distinction
between $\hat{\phi}_l^{\rm in}$ and $\hat{\phi}_l^{\rm up}$ disappears, so
that, if we demand that our solution to (\ref{de}) be both 
ingoing at the horizon and
outgoing at infinity, and then solve the resulting relation for $\omega$,
we know that we have found a pole.

\begin{figure}[h]
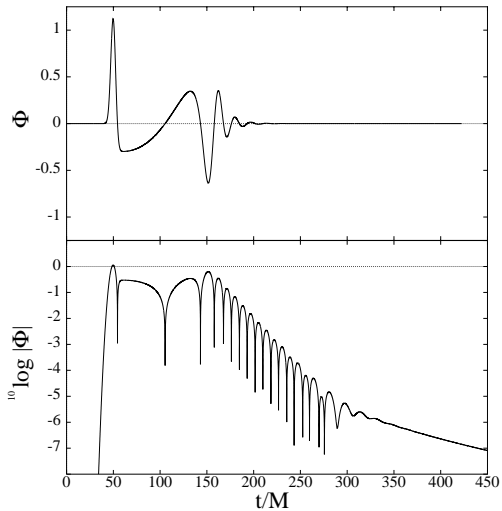

\hrule
\vishu
\hrule
\caption{A
recreation of Vishveshwara's classic scattering experiment: The response
of a Schwarzschild black hole as a Gaussian wavepacket of scalar
waves impinges upon it. 
The first bump (at $t=50M$) is the
initial Gaussian passing by the observer on its way towards the
black hole. Quasinormal-mode ringing clearly dominates the signal
after $t\approx 150M$. At very late times (after $t\approx 300M$)
the signal is dominated by a power-law fall-off with time.
This late time tail arises because of backscattering off of the 
weak potential in the far zone. As such, it is not an effect 
exclusive to black holes. A similiar tail will be present also for
perturbed stars. }
\label{vishu}
\end{figure}

It turns out that the corresponding frequencies, known as the quasinormal
modes of the black hole (quasinormal because they are damped as 
radiation dissipates to infinity and across the event horizon), 
play a dominant role in the evolution of black hole
perturbations.
The first indication of this was found 
by Vishveshwara~\cite{vishu}. He realized that  
one might be able to observe a solitary black hole by
scattering of radiation, provided the black hole left its fingerprint
on the scattered wave. So he started ``pelting'' the black hole with
Gaussian
wave packets. By tuning the width of the impinging Gaussian Vishveshwara
found that the black hole responded by ringing in 
a very characteristic decaying mode; the slowest damped of the 
black holes quasinormal modes. Subsequent work (in particular in 
numerical relativity) has shown that the quasinormal modes always 
play a prominent role in the dynamical response of a black hole to 
external perturbations (see Figure~\ref{vishu}). Impressive results for
head-on collisions  of two black holes lead to signals that are almost
entirely due to quasinormal mode ringing.

\begin{figure}[h]
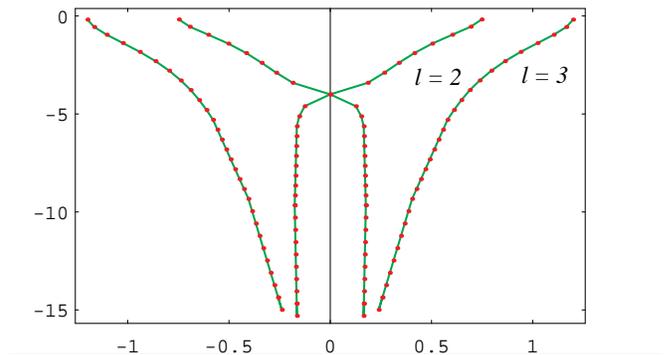

\caption{The complex quasinormal mode frequencies corresponding 
to gravitational perturbations (for $l=2$ and $3$)
of a Schwarzschild black hole.
These correspond to the positions of the poles of the radial Green's function
(equation \ref{greenf}) in
the complex $2M \omega$ plane.}
\hrule
\leaver
 \hrule
 \label{leaver}
 \end{figure}

The actual determination of  quasinormal mode frequencies
is a far from trivial calculation. 
The quasinormal modes are
solutions to (\ref{de}) that  satisfy the causal condition of
purely ingoing waves crossing the event horizon, while at the same
time behaving as purely outgoing waves reaching spatial infinity.
Assuming a time-dependence $e^{-i\omega t}$, a
general causal solution to (\ref{de}) is given by (\ref{inmode})
and a quasinormal mode corresponds to $A_{\rm
in}=0$. To identify a mode-solution we must therefore be able to
determine a solution that behaves as $e^{i\omega r_\ast}$ as
$r_\ast \to \infty$, with no admixture of ingoing waves. 
Assuming that the black hole is
stable (we can in fact prove that this must be the case), 
 no unstable mode-solutions should exist so we must require
that a mode is damped according to an observer at a fixed
location. This means that
$\mbox{Im }\omega_q<0$. The general solution (\ref{inmode}) 
is then a mixture of exponentially growing and  dying
terms. We must, out of all solutions, identify the unique 
one for which the coefficient of the exponentially
dying solution is zero.  
 Several methods have been devised to deal with
this difficulty accurately~\cite{nollert}. These methods
have been used to investigate the entire  spectrum for
non-rotating black holes, and also to map out the behaviour 
of the first ten modes or so as the black hole spins up.
The spectrum of gravitational-wave 
modes of a Schwarzschild black hole is shown in 
Figure~\ref{leaver}.

It is worthwhile to outline one of the most reliable methods
(due to Leaver~\cite{leaver}) for 
calculating black hole quasinormal modes.
Write the desired 
solution to equation (\ref{de}) as an infinite sum
\begin{equation}  
\hat{\phi}_l^{\rm in}  = (r-2M)^\rho (2M/r)^{2\rho} e^{-\rho(r-2M)/2M} \sum_{n=0}^\infty a_n 
\, \left({r-2M\over r}\right)^n
\end{equation}
where $\rho= -i2M\omega$. The recurrence relations between the 
$a_n$ are given by Leaver:
\begin{equation}
\alpha_n a_{n+1} + \beta_n a_n + \gamma_n a_{n-1} = 0,
\label{rec}
\end{equation}
where
\begin{eqnarray*}
\alpha_n &=& n^2 + 2n (\rho+1) + 2\rho+1 \\
\beta_n &=& -[2n^2 + 2n(4\rho+1)+8\rho^2 +4\rho+l(l+1) -s^2+1]\\
\gamma_n &=& n^2 +4 n\rho + 4\rho^2 -s^2  
\end{eqnarray*}
and $s$ is the spin of the field.
Now one can note that the coefficient $A_{\rm in}$ has a zero
whenever the  sum $\sum a_n$ converges.
This requirement translates into an continued-fraction equation involving 
the coefficients $\alpha$, $\beta$ and $\gamma$:
\begin{eqnarray}
&&\left[ \beta_q - {\alpha_{q-1} \gamma_q\over \beta_{q-1}-}
{\alpha_{q-2} \gamma_{q-1}\over \beta_{q-2}-}
\cdots
 {\alpha_{0} \gamma_{1} \over \beta_{0}}\right]
\\ \nonumber
&\quad &\qquad\qquad =
\left[ {\alpha_{q} \gamma_{q+1}\over \beta_{q+1}-}
{\alpha_{q+1} \gamma_{q+2}\over \beta_{q+2}-}
{\alpha_{q+2} \gamma_{q+3}\over \beta_{q+3}-}
\cdots
\right].
\label{contfrac}
\end{eqnarray}
Here $q=1,2,3,\dots$ is the mode number. Solving
this equation numerically (still a non-trivial task!) 
for $\omega_q$ gives the quasinormal modes (see Figure \ref{leaver}).

\subsection{Mode excitation}

Having located the quasinormal modes we want to 
evaluate the contribution of each mode to the emerging signal.
Ideally one would like to be able to quantitatively account for the
contribution to a signal from each individual 
quasinormal mode. Thus we want to construct the 
mode-contribution to the Green's function (\ref{greent}),
combine it with the relevant initial data an extract the 
corresponding signal using (\ref{Gevol}). To do this is largely
a (rather involved) numerical exercise. In the end one finds that the
quasinormal modes account for the main part of the signal after a
certain time, essentially the time it takes for a part of the initial
data to travel from its original position $y$ to the black hole, and then 
for the scattered wave to reach the 
observer  at $r_\ast$. Thus we expect the modes
to generally dominate for (roughly) $t-r_\ast-y>0$. 
 
It is helpful to introduce a simplifying approximation at this stage.
Let us assume that the initial data has support only far away from the black 
hole, and that the observer is also located in the far zone. Then we can can replace the solutions $\hat{\phi}_l$ in (\ref{greenf}) by their asymptotic
behaviour at large $r_\ast$ and readily evaluate the mode 
contribution to the Green's function. 
Since  $A_{\rm in}(\omega)  $ has a simple zero at 
$\omega= \omega_q$,
  it is useful to
define a quantity $\alpha_q$ by
\begin{equation}
A_{\rm in}(\omega) \approx (\omega -\omega_q) \alpha_q \ ,
\label{alph}\end{equation}
in the vicinity of the pole. 
Then it follows from the residue theorem (and the fact
that modes in the third and fourth quadrant are in one-to-one
correspondence, see Figure~\ref{leaver}) 
that the total contribution from the 
modes to the time-domain Green's function can be written 
\begin{equation}
G^Q(r_\ast,y,t) =  {\rm Re} \left[ \sum_{q=0}^\infty 
B_q e^{-i\omega_q ( t-r_\ast-y)} 
\right] \ .
\label{Gmodes}\end{equation}
Here we have defined
\begin{equation}
B_q = {A_{\rm out} (\omega_q)\over \omega_q \alpha_q }  \ .
\end{equation}
The sum in (\ref{Gmodes}) is over all quasinormal modes 
in the fourth quadrant of the $\omega$-plane. That
this expression provides an accurate representation of the
mode-excitation (as long as our assumptions are valid)
has been demonstrated.
 
At this point it is relevant to comment on the fact that the 
quasinormal modes are, even though there is an infinite set of
modes for each $l$, not complete. That is, a mode sum such as 
(\ref{Gmodes}) should not be expected to represent the entire
black-hole signal for given initial data. It will typically not 
be useful at early times, and it cannot represent the
power-law tail that dominates at very late times, see 
Figure~\ref{vishu}. However, the mode-contribution is 
still highly relevant, and results like (\ref{Gmodes})
can help us understand the dynamics of black holes better. 
Furthermore, expression (\ref{Gmodes}) allows us to study the
convergence of the mode-sum in a simple way. It has been shown that
(again under the assumptions of the ``asymptotic approximation'')  
that the mode sum converges for  $t-r_\ast-y>0$. 

\subsection{Useful approximations}

The quasinormal modes provide (at least in principle) a unique
way of identifying black holes and deducing their mass and rate of rotation.
Given this, it instructive to have simple approximations of the most
important mode-frequencies. We can readily arrive at such expressions 
by recalling that the black hole problem is essentially one of scattering off
a single potential peak (close to $r=3M$).  
It is commonly
accepted that scattering resonances (the
quantum analogues of quasinormal modes) arise for energies close to the
top of a potential barrier. 
This immediately leads to the approximation
\begin{equation}
{\rm Re}\ \omega_0 
\approx {1 \over 3 \sqrt{3}M} \left( l+{1\over 2} \right) \ .
\label{reapp}\end{equation}
This is a good approximation of the fundamental
Schwarzschild quasinormal (gravitational wave)
mode for large $l$.
For the imaginary part of the
frequency---in quantum language: the lifetime of the 
resonance---the curvature of the
potential at the peak contains the relevant information. 
Schutz and Will~\cite{wkb} used the WKB approximation to infer that
\begin{equation}
{\rm Im}\ \omega_0 \approx - {\sqrt{3} \over 18 M} \ ,
\label{imapp}\end{equation}
which is accurate to within 10 percent for the fundamental mode.

Let us translate the results for the fundamental gravitational-wave
quasinormal mode of a nonrotating black hole into more familiar units. 
We then get a frequency
\begin{equation}
f \approx 12 {\rm kHz } \left( {M_\odot \over M} \right) , 
\end{equation}
where $M_\odot$ represents the mass of the Sun,
while the associated e-folding time is
\begin{equation}
\tau \approx 0.05\ {\rm ms } \left( {M \over M_\odot} \right) \ .
\end{equation}
The quasinormal modes of a black hole are clearly very shortlived.  
In fact, we can 
compare a black hole to
other resonant systems in nature by considering the quality factor 
\begin{equation} 
Q \approx {1 \over 2} \left| { \mbox{Re }\omega_q \over \mbox{Im }\omega_q } \right| \ .
\end{equation}
Our 
quasinormal-mode approximations then lead to $Q\approx l$. This should be
compared to the result for the fluid pulsations of a neutron star:
$Q\sim 1000$, or the typical value for an atom: $Q\sim 10^6$. A 
Schwarzschild 
black hole is clearly a very poor oscillator.

\section{Complex angular momentum approach}

As early as 1972 there was an intriguing suggestion (due to 
Goebel\cite{goebel}), 
that the 
black-hole resonant modes could have the following physical interpretation: 
a standing wave could establish itself along the stable photon 
orbit at or near radius $r=3M$.  Of course this standing wave is not 
stable but  would  decay by radiating away energy.
Unlike the quasi-normal modes, Goebel's standing waves correspond to poles 
of the Green function in the
complex {\sl angular momentum} plane. 
Thus we need to extend our previous analysis to 
allow for complex values of $l$. That this leads to a powerful
description of many scattering problems is well known~\cite{nuss}. 
The complex angular momentum paradigm has been  
much investigated
in acoustical and electromagnetic scattering, but  
has received scant attention in the context of black holes.

\subsection{Cross sections}

\begin{figure}[h]
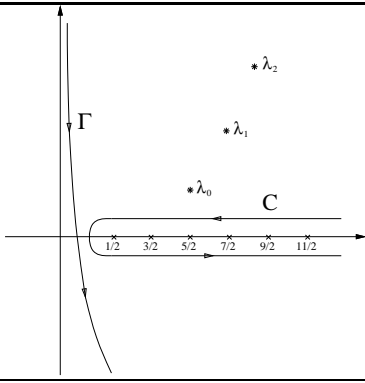

\caption{Integration contours in the complex $\lambda (=l +1/2) $-plane used
in the derivation of the CAM representation for the scattering
amplitude, equation (\ref{cscamp}). 
$C$ is the original contour used in the integral
representation for the scattering amplitude. 
The relevant contour for the background integral
in the CAM picture is $\Gamma$. 
The Regge poles $\lambda_n$ are all situated in the
first quadrant. Their contribution is accounted for by the
residue-theorem.} 
 \hrule
\cont2
 \hrule
\label{cont2}
\end{figure}

Let us begin by reviewing the theory of complex angular-momentum (CAM)
scattering. Let $F(l+1/2)$ be any function which is analytic in the
neighbourhood of the positive real $l$ axis. Then, by Cauchy's theorem,
we may write:
$$
\sum_{l=0}^\infty (-1)^l F(\lambda) = 
{i\over 2} \oint_C {F(\lambda) \over \cos \pi \lambda} \ d\lambda
$$
where $\lambda = l+  1/2$. We apply this transformation to equation 
(\ref{scamp}),
writing
$$
f(\theta) = {1\over 2 \omega} \oint_C {\lambda [S_l -1] P_l (-\cos\theta) 
\over \cos \pi \lambda} \ d\lambda .
$$
We deform the contour $C$ away from the real $l$ axis, rewriting $f$ as
the sum over the poles in $S_l$ and a background integral
(see Figure~\ref{cont2}).
\begin{eqnarray}
f(\theta) &=& f_P + f_B = {-i \pi \over \omega} \sum_n 
{\lambda_n r_n \over \cos \lambda_n} \nonumber
\\
&& + {1 \over 2\omega} \int_\Gamma {\lambda [S_l -1] P_l (-\cos\theta) 
\over \cos \pi \lambda} \ d\lambda .
\label{cscamp}
\end{eqnarray}
Here $r_n$ is the residue associated with the pole $l_n$, defined
by 
\begin{equation}
S_l \approx {r_n \over l - l_n} \ ,
\label{resis}\end{equation}
in the vicinity of the $n$th pole. Even though we will not consider
the `background integral' further in this article, it is worth
mentioning that it can be approximated using the saddle point method.

\begin{figure}[ht]
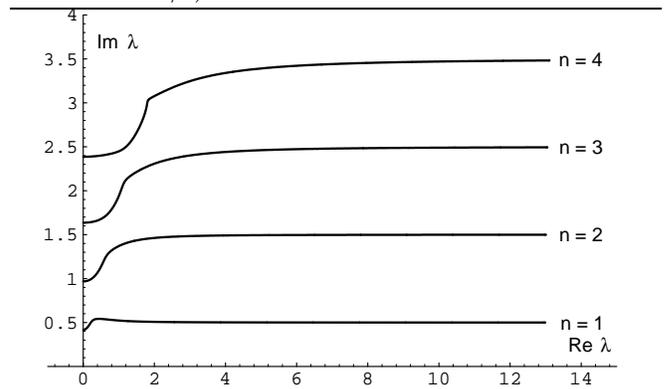

\caption{The trajectory of the Regge poles $l_n(\omega)$ ($n=1,2,3,4$)
followed in the complex $l$ plane
from  $\omega M= 0$ (on the line $l=-1/2$) to $\omega M = 5$.}
\hrule
\suivi2
\smallskip
\hrule
\label{suivi2}
\end{figure}

The poles of $S(\lambda)$ are known as Regge poles and
for the black hole potential one
can show (Andersson and Thylwe, 1994)
that they are all located
in the first quadrant of the complex $\lambda$-plane.
If
If we compare the solution for one of the 
Regge poles
 to that of a quasinormal
mode we see that they are rather similar. Both solutions satisfy
purely ``outgoing'' wave boundary conditions. 
Hence, methods used for finding the quasi-normal mode frequencies can
be adapted to find the Regge poles.
One can show that these poles are all located in the first quadrant of
the complex $\lambda$-plane.  
Typical results for the first few poles
are shown in Figure~\ref{suivi2}.

\section{Physical interpretations}

An important aspect of the CAM description of scattering is that each
Regge pole  has a clear
interpretation. To realize this two approximations for the Legendre
functions are useful:

For $\theta$ not close to $0$ and $\vert \lambda_n \vert >>1$ we 
can use the formula 
\begin{equation}
P_{l_n} (-\cos \theta) \approx \left( {\pi - 
\theta \over \sin \theta} \right)^{1/2} J_0 (\lambda_n 
(\pi - \theta)) \ ,
\label{glory1}\end{equation}
when evaluating the Regge-pole sum (\ref{cscamp}).  This is especially
interesting since we know that black-hole cross sections show a
prominent glory in the backward direction, see Figure~\ref{cross}. 
It is commonly understood that glories
are characterized by Bessel-function type oscillations, cf. (\ref{glapp}).
When numerical results for the Regge pole with the smallest 
imaginary part (for a given $\omega$) is used in (\ref{glory1})
we get a good approximation to the glory oscillations in the black-hole
cross section.

Alternatively, we can use an asymptotic approximation for the Bessel 
function in (\ref{glory1}). We then get 
\begin{eqnarray}
P_{l_n}(-\cos \theta) 
\approx  { e^{i\lambda_n(\pi-\theta) - i\pi/4} +
e^{-i\lambda_n(\pi-\theta) + i\pi/4} \over \sqrt{ 2\pi \lambda_n \sin
\theta} } 
\label{surface}\end{eqnarray}
for $\vert \lambda_n\sin\theta \vert \to \infty $. From this formula it
follows that we may interpret each Regge state as a combination of two
surface waves travelling around the scattering centre (the black hole)
in opposite directions. The angular velocity of each wave is
proportional to $1/{\rm Re}\ \lambda_n$, and as they propagate around
the black hole the waves decay exponentially.  The imaginary part of
$\lambda_n$ is clearly associated with the inverse of the ``angular
life'' of each surface wave. It also follows, since ${\rm
Im}\ \lambda_n>0$, that the amplitude of the first term in
(\ref{surface}) is, in general, smaller that the second term. Only for
$\theta \approx \pi$ do the two amplitudes have similar magnitude.
Hence, one would expect interference effects to be more pronounced in
the backward direction. Moreover, it is easy
to show that the anticipated diffraction oscillations will have a
period of $\pi/{\rm Re}\ \lambda_n$. Again, this result approximates
the features seen in the black-hole cross section rather well.

\begin{table}[h]
\caption{``Angular life'' and impact radius ($R_n$) for the first few
 Regge poles for $\omega M = 2$. It should be remembered that the impact parameter
associated with the unstable photon orbit at $r=3M$ is roughly $5.196
M$.
}
\begin{tabular}[t]{lcc}
 $n$  & Impact radius ($M$) &  Angular life (degrees)\\
\hline
0& 5.194 & 114 \\
 1 & 5.207 & 38 \\
 2 & 5.234 & 23
\end{tabular}
\label{angtab}
\end{table}

We can also use the standard localization principle (cf. the definition
of the impact parameter)
\begin{equation}
{\rm Re}\ \lambda_n \approx \omega R_n \ ,
\end{equation}
to infer that the real part of each pole position is associated with
the distance from the black hole at which the angular decay occurs. In
the case of a Schwarzschild black hole one would expect such surface
waves to be localized close to the unstable photon orbit at $r=3M$ [or,
strictly speaking, the maximum of the effective potential in
(\ref{de})]. This would correspond to $R_n = 3 \sqrt{3} M \approx
5.196 M$. As
can be seen from Table~\ref{angtab} 
the first Regge pole for various frequencies
leads to a value of $R_n$ that is close to the impact
parameter for the photon orbit. 
 
\subsection{Sample results}

\begin{figure}[h]
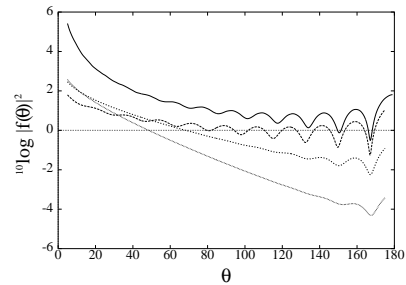

\caption
{ The differential cross section for $\omega M = 2.0$ as obtained from
the phase-shifts (solid line) is compared to the
contribution from each of the first three Regge poles (dashed lines). }
 \hrule
\rcross
 \hrule
\label{reggecross}
\end{figure}

Although the physical interpretations discussed above are
suggestive and agree well with the partial-wave results for the cross 
section we should also compute the actual cross sections before
assessing the usefulness of the CAM approach to black-hole scattering.
In Figure~\ref{reggecross} we show the 
contribution to the the pole-sum $f_P$ from the
first three Regge poles for $\omega M=2.0$. The results are compared to
the partial-wave cross section as computed from the partial-wave sum
(Figure~\ref{cross}). To obtain the Regge-pole contributions we have used
the asymptotic formula (\ref{surface}) for the Legendre functions.

From the data shown in Figure~\ref{reggecross}
we make two observations: 
i) For large scattering angles ($\theta \ge 40^\circ$) the pole sum
in~(\ref{cscamp}) is dominated by the Regge pole with the smallest
imaginary part. Each consecutive pole gives a contribution that is
roughly two orders of magnitude smaller than that of the preceding
pole. This means that only one Regge pole need be included in a
reasonably accurate description of the black-hole glory.
ii) For smaller angles ($\theta \le 40^\circ$) the first three terms
in the pole sum are of the same order of magnitude. This is consistent
with the interpretation of $1/{\rm Im}\ \lambda_n$ as the ``angular
life'', see Table~\ref{angtab}.

In conclusion, we have shown that the Regge states can be 
interpreted as surface
waves that travel around the black hole. At the same time the waves
decay at a rate related to the imaginary part of the Regge pole
position. We have also seen that the glory oscillations that arise for
large scattering angles in the black-hole case are naturally described
in the CAM representation.  In the specific example presented 
(for $\omega M=2.0$) a single
 Regge pole accounts for all large-angle structure in the scattering
 cross section. 


\section{Quantum effects}

We will now turn to another extreme scale in physics, the level 
where the classical theory of general relativity breaks down and
it must somehow be married to quantum theory. We will not attempt
to describe the
ongoing attempts to formulate theory of quantum gravity in detail.
Rather, we want to point out that 
 one can gain some insights into quantum gravity
(at the semiclassical level) using concepts and techniques 
very similar to those we have discussed above. In fact, 
several interesting effects 
bear a great resemblance to various scattering
scenarios. Since we will only skim the surface of a vast area
of research, we refer the reader to the monograph by 
Frolov and Novikov~\cite{frolov} for references and
more details.

\subsection{Hawking radiation}

\begin{figure}
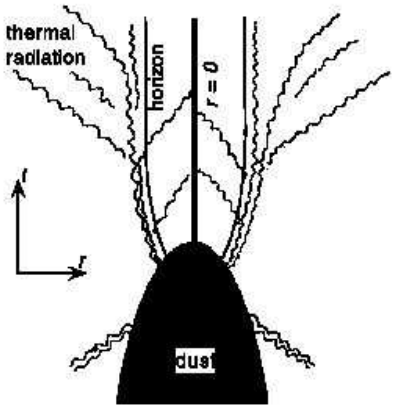

\caption
{The Hawking effect: when dust collapses to form a black hole, 
high-frequency vacuum fluctuations near the horizon lead to 
radiation which can escape to infinity.}
\hrule
\thorne
\hrule
\label{thor}
\end{figure}

Consider Figure~\ref{thor}. A cloud of pressureless dust undergoes collapse
to form a black hole. The wiggling lines represent high-frequency waves 
associated with a massless field (scalar, neutrino, photon, graviton).
We suppose that the space is initially indistinguishable from Minkowski
space and that the field is in the vacuum state. Now consider the evolution of
a high-frequency wave packet that is a part of the spectrum of (Minkowski)
vacuum fluctuations. We follow the evolution of the packet which,
after the collapse, is localised just outside the black hole's horizon
at $r=2M$.

The packet can be spectrally decomposed into a component which is ingoing
and another which (eventually) can escape to infinity.  Hawking~\cite{hawk} 
showed
that, when viewed from infinity,  the spectrum of the escaping radiation
is thermal with temperature $T_H = \hbar/(k_B 8\pi M)$. That is, 
if we Fourier-decompose the radiation escaping to infinity and
use the standard definition of the vacuum state, we find that the
total number of particles in the mode $(n)$ is 
$$
N_{(n)} = |T_{(n)}|^2/(e^{\hbar \omega/(k_B T_H)} \mp 1)
$$
Here $(n)$ is a general index which represents all quantum numbers
and the   sign is taken for negative for integer and
 positive for half-integer spin fields.

We see the connection with scattering theory when
we write the expression for the
total (scalar) luminosity of the black hole due
to Hawking radiation:
$$
L = {\hbar \over 2\pi} \int\limits_0^\infty {\omega \ d\omega \over
e^{\hbar \omega/k_B T_H} -1}  \sum_{l=0}^\infty (2l+1) |T_l(\omega)|^2.
$$
There are similar formulas for quantum fields of other spins. 
For a stellar-mass black hole, this is
$$
L_{\hbox{\small scalar}} = 7.44 \times 10^{-5} \hbar/M^2
$$
Reverting momentarily to cgs units, we can express the luminosity
for a field of spin $s$ as
$$
L_s  = \alpha_s \cdot 4.09 \times 10^{-17} 
\left(M_\odot \over M_{\hbox{{\small BH}}}\right)^2  \hbox{ergs/sec}.
$$
where $M_\odot$ is the mass of the sun, and the 
$\alpha_s$ are given in Table \ref{luminosities}. Clearly the Hawking effect
is too small to ever detect for a black hole of near stellar mass.
But the effect may be appreciable for much smaller black holes
conceivably created in the early universe. Such primordial black holes
would radiate at an appreciable level, and there have been 
speculations that one might be able to observe bursts of radiation
as these black holes evaporate.  As yet this
effect has, however, not been observed.

\begin{table}[h]
\begin{tabular}[t]{l|l|l}
field & spin $s$&  $\alpha_s = L M^2/\hbar$\\
\hline
scalar & 0 &$7.44 \times 10^{-5}$ \\
neutrino & $1/2$ & $8.18  \times 10^{-5}$ \\
photon & 1 & $3.37  \times 10^{-5}$ \\
graviton & 2 & $0.38 \times 10^{-5}$ \\
\end{tabular}
\caption{Black hole luminosity as a function of field spin $s$}
\label{luminosities}
\end{table}

\def\diag{{\hbox{\rm diag}}}
\def\tmunu{{\langle {\hat T}_\mu{}^\nu \rangle}}
\def\Tloc{{T_{\hbox{{\small loc}}}}}
\subsection{Stress-energy tensors}

It must be emphasised that the Hawking radiation emitted by
the black hole is not produced by the process of collapse
but is feature of the  field vacuum state in the new spacetime,
which now contains a black hole. One of the lessons of quantum field
theory in curved spacetime is that we must mistrust the abstract concept 
of `particle' and concentrate on physically measurable quantities such
as the energy content of the field in a particular state, 
that is, the stress-energy tensor of the
quantum field, written $\tmunu$. It turns out that one can 
determine the stress-energy tensor via a differential operator
acting on the black-hole perturbation Green's function that 
we have introduced earlier.  

It will be  useful for what follows to define the
{\em Tolman local temperature}
$$
\Tloc = T_H g_{00}^{-1/2} = T_H/\sqrt{1-2M/r} .
$$
A massless scalar field at temperature $T$ has  
$$
\tmunu_\flat ={\pi^2 \over 90} T^4 \diag(-3,1,1,1)
$$
in flat space.
A gas of scalar particles in equilibrium 
exterior to a spherical body should have
$$
\tmunu ={\pi^2 \over 90} T_{\hbox{{\small loc}}}^4 \diag(-3,1,1,1)
$$
that is, $T$ is replaced by the local temperature.
This expression diverges as $r \to 2M$.

The most obvious choice for the vacuum state exterior to
a black hole is the state that, at infinity, resembles
as much as possible the Minkowski vacuum state. This state,
called the Boulware state, is the vacuum state appropriate for
a field outside of a spherical body which is larger than its
Schwarzschild radius, such as a neutron star. When $\tmunu$ is
calculated in this state it has the asymptotic form
$$
\tmunu_B \sim -{\pi^2 \over 90} T_{\hbox{{\small loc}}}^4 \diag(-3,1,1,1) 
\qquad r \to 2M.
$$
In other words, the energy density is negative infinite on {\bf H}$^+$ and
{\bf H}$^-$, cf. Figure~\ref{inup}. 
Note that this infinity has nothing to do with renormalisation:
the divergences in $\tmunu$ have already been removed using covariant,
state-independent methods.
The presence of a singularity in the stress-energy tensor, even when measured by
a freely-falling observer, is unphysical and we conclude that the Boulware state
cannot be the ground state of a field exterior to a black hole.

If we are interested in a black hole formed by stellar collapse, we would
like the stress-energy tensor to (at least) be regular on {\bf H}$^+$
and to tend to the Minkowski vacuum tensor on {\bf I}$^-$. The price we
pay for this requirement is that the tensor is singular on {\bf H}$^-$.
(Of course, in the case of stellar collapse, {\bf H}$^-$ does not exist.)
In addition, at {\bf I}$^+$ we have an outflow of thermal radiation 
as described in the preceding section. This state is called the
Unruh state.

If we demand that the stress-energy tensor be regular on both
{\bf H}$^-$ and {\bf H}$^+$
 the Hartle-Hawking state, and is appropriate for a
spacetime with an eternal black hole in equilibrium with a thermal bath
of radiation
at the Hawking temperature. The stress tensor has the form:
\begin{eqnarray*}
\tmunu_H &=&  \\
 &&{\pi^2 \over 90} T_{\hbox{{\small loc}}}^4 \diag(-3,1,1,1) 
 \left[1 - \left({2M\over r}\right)^6 \left(4 - {6M \over r}\right)^2\right]
 \\
&&\qquad  + \hbox{finite} 
\end{eqnarray*}
where `finite' is a correction everywhere finite and of order $r^{-6}$ for
large $r$. Despite its appearence, $\tmunu_H$ is
finite at $r=2M$. 
  
We see then that the three `vacuum' states correspond to different
physical situations, and that it is not possible to define a state
which has the properties which we naturally associate with the vacuum.
Although we rejected the Boulware state because of infinite
energy on the horizon, the Hartle-Hawking state also has infinite
energy, which is contained in the heat bath at infinity. 
The Unruh state, the most `realistic' state,
 is appropriate for an eternally evaporating black hole of constant mass.
A real black hole will lose mass  by evaporation, evolving to
an unknown final state. 
This unsatisfactory
situation will only be resolved by a full quantum theory of gravity.

\section{The Kerr Black Hole} 

So far we have only considered the simplest class of black holes, namely
those described by the spherically symmetric Schwarzschild 
solution to Einstein's equations. There are other kinds of black holes
as well. In order to generalise our discussion to the case of 
greatest physical interest we must allow our black hole to rotate. 
 Due to conservation of angular momentum 
during the gravitational collapse one would expect most newly
born black holes to 
spin very fast.
Then the resultant spacetime metric will no longer be spherically
symmetric. The corresponding solution to Einstein's equations was 
discovered by Kerr, and the metric is usually written
\begin{eqnarray}
ds^2
& = &
-{\Delta \over \rho^2 } [dt - a \sin^2\theta d\varphi]^2 + {\sin^2\theta \over \rho^2}
[(r^2+a^2)d\varphi - a\, dt]^2 \nonumber \\
&&\qquad\qquad + {\rho^2 \over \Delta} dr^2  + \rho^2 d\theta^2 
\end{eqnarray}
\index{black hole!rotating}
where
\[
\Delta= r^2 - 2Mr + a^2, \qquad \rho^2 = r^2 + a^2 \cos^2 \theta
\]
Here $a$ is a parameter representing the angular momentum per unit mass.
When $a$  equals zero the Kerr metric reduces to the Schwarzschild metric.
The event horizon of a rotating black hole corresponds to the 
outer solution to $\Delta = 0$, and is given by
\begin{equation}
r_+ = M + \sqrt{M^2-a^2} \le 2M.
\end{equation}

In studying scattering from rotating black holes all the concepts
we have introduced for non-rotating black holes remain useful.
Thus we only need to comment on how these results are affected
by the black holes rotation. It is natural to begin by discussing the 
nature of light trajectories in the Kerr spacetime.

\subsection{Null geodesics in the Kerr geometry}

A description of the general trajectories of a particle moving in the
Kerr geometry is considerably more complicated than the Schwarzschild
case, but because of the axial symmetry one would still expect  
$p_\theta=0$ for motion in the equatorial plane ($p_\mu$ 
represents the four-momentum of a photon). If a 
particle is initially moving in the equatorial plane, it should remain
there.
In the Schwarzschild case we could always, because of the spherical 
symmetry, orient our coordinate system in such a way that a study of 
equatorial trajectories covered all possible cases. For
rotating black holes, we no longer do this and an equatorial
trajectory is a very special case. Nevertheless, they provide a
useful starting point for an exploration of particle motion around a
rotating black hole. 

Because  the Kerr geometry is stationary and axisymmetric, 
we will still have the
two constants of motion $p_t = -E$ and $p_\varphi= L_z$,  
the ``energy measured at infinity'' and the component of the angular
momentum parallell to the symmetry axis of the spacetime. 
Given this we can immediately deduce two equations of motion
\begin{eqnarray}
p^t &=& {dt \over d\lambda} = { (r^2+a^2)^2-a^2\Delta\sin^2 \theta \over
\Sigma \Delta} E - {2aMr \over \Sigma \Delta}L_z \ , \label{eomt} \\
p^\varphi &=&  {d\varphi \over d\lambda} = {2aMr \over \Sigma \Delta}E
+ { \Delta -a^2\sin^2 \theta \over \Sigma\Delta\sin^2 \theta} L_z \ ,
\label{eomphi}\end{eqnarray}
where $\Sigma^2 = (r^2 + a^2)^2 - a^2 \Delta\sin^2\theta$. 

We will
restrict our attention to photons moving in the equatorial plane. Then
the above formulae together with $p_\mu p^\mu =0$ and $p_\theta = 0$ 
lead to an equation
for the radial motion that can be factorized as
\begin{equation}
\left( {dr \over d\lambda} \right)^2 = {(r^2+a^2)^2 - a^2 \Delta \over
r^4}
(E-V_+)(E-V_-) \ , 
\label{radkerr}\end{equation}
where we have defined
\begin{equation}
V_\pm(r) = {2aMr \pm r^2 \Delta^{1/2} \over (r^2+a^2)^2 - a^2 \Delta}
L_z \ .
\label{vpmkerr}\end{equation}
By expanding these potentials in inverse powers of $r$ we see that
they fall  off as $1/r$ as
$r\to\infty$ (as in the Schwarzschild case), but the effect
of rotation enters at order $r^{-3}$. 
This means that the rotation of the black hole has
little effect on a distant photon. But as the photon approaches the
black hole the potentials have a much stronger influence and we can 
distinguish two different cases. The way that the rotation of the
black 
hole
affects an incoming photon depends on direction of  
$L_z$ relative to the sense of
rotation of the black hole. 

\begin{figure}[h]
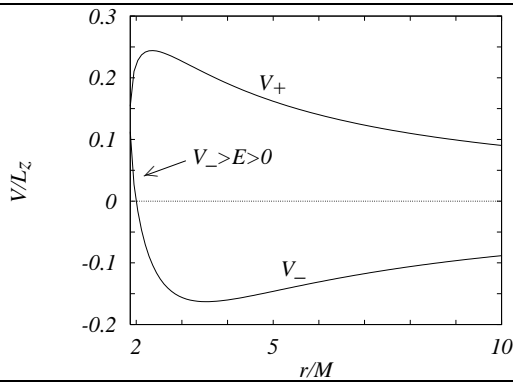

\caption{The effective potentials for a photon moving in the
equatorial plane of a rotating black hole. The figure illustrates
the
case when the photon has angular momentum directed in the same sense
as the rotation of the hole ($a=0.5M$). The corresponding figure for a
retrograde photon (with $aL_z<0$) is obtained by turning this figure 
upside-down. 
} 
\hrule
\vkerr
\hrule
\label{vkerr}
\end{figure}

In the case when $aL_z>0$, when the photon moves around the black hole
in a prograde orbit,  we get
the
situation illustrated in Figure~\ref{vkerr}.
Then we can see from (\ref{vpmkerr}) that 
\begin{eqnarray}
V_- &=& 0 \mbox{ at } r = 2M \ , \\
V_+ &=& V_- = { aL_z \over 2Mr_+} = \omega_+ L_z   \mbox{ at } r =r_+\ ,
\end{eqnarray}    
where we have defined the angular velocity of the horizon $\omega_+$.

Since the lefthand side of (\ref{radkerr}) must be positive (or zero)
we can
infer that the photon must either  move in the region $E>V_+$ or in $E<V_-$.
In the first case the result is familar. 
An incoming photon from infinity can either be scattered by, or
plunge into, the black hole. 
But what about the region $V_->E>0$ that would also 
seem to be accessible (cf. Figure~\ref{vkerr})?
An analysis of this possibility requires some care. It turns out that
it is not sufficient to require that $E>0$, as one might intuitively
think.
The reason for this is quite easy to
understand: $E$ is the energy measured at infinity, and as we get
closer to the black hole it becomes a less useful measure of what is
going on. In order to understand the properties of light trajectories
in the vicinity of a spinning black hole, 
we need an observer located close to the horizon to do our
measurements for us. 

A convenient choice of local observer is one that has zero-angular
momentum
and resides at a fixed distance from the black hole (at constant $r$).
Such Zero-Angular Momentum Observers (ZAMOs) were first introduced by
Bardeen~\cite{bardeen}. (It should be noted that a ZAMO does not follow a geodesic,
and 
consequently must maintain its position, say, by means of a
rocket.)
The suggested character of a ZAMO means that it must have a four-velocity
\begin{equation}
u^t = A \ , \qquad u^\varphi = \omega A , \qquad u^r = u^\theta =0 \ .
\end{equation}  
The unknown coefficient $A$ is specified by the requirement
\begin{equation}
u_\mu u^\mu = -1 \ ,
\end{equation} 
and we find that
\begin{equation}
A^2 = { g_{\varphi \varphi} \over (g_{\varphi t})^2 - g_{tt}g_{\varphi
\varphi} } \ . 
\end{equation} 

We are now equipped to address the question of photons in the region
$V_->E>0$ in Figure~\ref{vkerr}. A ZAMO will measure the energy of
a photon as
\begin{equation}
E_{\rm zamo} = -p_\mu u^\mu = - (p_t u^t + p_\varphi u^\varphi) =
A(E-\omega L_z) \ .
\end{equation}
This ``locally measured'' energy is the one that we
must require to be positive
definite,
which means that we must have $E>V_+$ in Figure~\ref{vkerr}.
In other words, the possibility $V_->E>0$ is not physically acceptable
and we can conclude that the case $aL_z>0$ only contains the same 
types of photon trajectories as we found in the Schwarzschild case.

This is not, however, true for the case when $aL_z<0$, when 
the photon  is inserted in a
retrograde orbit around the black hole. (The potentials for this case
are easily obtained by turning the ones in Figure~\ref{vkerr} upside down.)
We then find from (\ref{vpmkerr}) that  
\begin{equation}
V_+ = 0 \mbox{ at } r = 2M \ , 
\end{equation}  
and it is clear that some forward-going photons (that must lie above
$V_+$ according to our previous analysis) can have $E<0$. That is,
negative
energy (as measured at infinity) photons may exist close to the black
hole
(for $r<2M$ in the equatorial plane)! 
As can be inferred from 
Figure~\ref{vkerr} these negative energy photons can never
escape
to infinity, but the fact that they can  exist  has an interesting  
consequence. 

Let us suppose that a pair of photons, the total energy
of which is zero, 
 are created in the region $r_+<r<2M$. The positive energy photon can
then escape to infinity, while the negative energy one must eventually
be swallowed by the hole. The net effect of this would be that
rotational
energy is carried away from the black hole, and it will slow down.
This energy extraction process, that was first suggested by 
Penrose~\cite{penrose}, 
can be extended to other objects. One can simply assume that a body
breaks up into two or more pieces. If one of them is injected into a
negative energy orbit the sum of the total energy of the remaining
pieces must be greater than  the total energy of the original
body, 
since $E$ is a conserved quantity. As in the case of photons, 
the extra energy is mined from the
rotation of the black hole.  
But however exciting the possibility may seem, the Penrose process 
is unlikely  to play a role in an astrophysical setting.      

\subsection{The ergosphere}

As we have seen, there is a region close to a rotating black hole
($r<2M$ in the equatorial plane) where energy becomes `peculiar'.
This is the so-called ergosphere, and since there are
many interesting effects (like the Penrose process)
associated with it, it is worthwhile to discuss it in more detail.

Let us consider a photon emitted at some $r$ in the equatorial
plane ($\theta=\pi/2$) of a Kerr black hole. Assume that the photon 
is initially moving in the $\pm \varphi$ direction. That is, it is
inserted in an orbit that is tangent to a circle of constant $r$. 
In this situation it is clear that only $dt$ and $d\varphi$ will be
nonzero, and
we find from $ds^2=0$ that
\begin{equation}
{d\varphi \over dt} = -{g_{t\varphi} \over g_{\varphi\varphi}} \pm
\sqrt{ \left( {g_{t\varphi} \over g_{\varphi\varphi}} \right)^2 - 
{g_{tt} \over g_{\varphi\varphi}} } \ . 
\end{equation}      
From this we can see that something interesting happens if $g_{tt}$
changes sign. At a point where $g_{tt}=0$  we have the two solutions
\begin{equation}
{d\varphi \over dt} = -2  {g_{t\varphi} \over g_{\varphi\varphi}}  \ ,
\qquad \mbox{ and } \qquad  {d\varphi \over dt} = 0 \ .
\end{equation}
In the Kerr geometry the first case corresponds to a photon moving in
the direction of the rotation of the black hole. The second solution,
however, indicates that a photon sent ``backwards'' does not
(initially) move at all! The dragging of
inertial frames has become so strong that the photon cannot move in
the
direction opposite to the rotation. 
Consequently, all particles must rotate
with the hole, and no observers can remain at rest (at constant
$r,\theta,\varphi$) in the ergosphere.
 
As the above example indicates, the boundary of the ergosphere follows from
$g_{tt}=0$. In the Kerr case we find that this corresponds to
\begin{equation}
\Delta - a^2\sin^2\theta = 0 \ ,
\end{equation}
or
\begin{equation}
r_{\rm ergo} = M \pm \sqrt{ M^2-a^2\cos^2\theta} \ .
\end{equation}      
From this  follows that the ergosphere always
lies outside the event horizon (even though it touches the horizon 
at the poles).  

\subsection{Teukolsky's equation}

We want to extend our study of various scattering scenarios
to the Kerr case. To do so, we need to discuss perturbations of
the Kerr geometry. It turns out that this problem is considerably
more complicated than the Schwarzschild one. For example,  
 the direct derivation of the equations governing
perturbations of Kerr spacetimes by considering perturbations of
the metric fails. It leads to gauge-dependent, and rather messy, formulations
in which one cannot readily separate the variables as in (\ref{sdec}).

A theoretically attractive alternative is to
examine {\em curvature} perturbations.  Using the
Newman-Penrose formalism, Teukolsky (1973) 
derived a master equation
governing not only gravitational perturbations (spin weight $s = \pm 2$) but
scalar ($s=0$), two-component neutrino ($s=\pm 1/2$) and electromagnetic 
($s=\pm 1$) fields as well.
In Boyer-Lindquist coordinates and with
the use of the Kinnersley null tetrad, this master
evolution equation reads
\begin{eqnarray}
\label{teuk0}
&&
-\left[{(r^2 + a^2)^2\over \Delta} -a^2\sin^2\theta\right]
         \partial_{tt}\Psi
- {4 M a r \over \Delta}
         \partial_{t\phi}\Psi
\nonumber\\   &&
   - 2s\left[r- {M(r^2-a^2) \over \Delta} +ia\cos\theta\right]
         \partial_t\Psi \nonumber\\   &&
+\,\Delta^{-s}\partial_r\left(\Delta^{s+1}\partial_r\Psi\right)
 +  {1\over \sin\theta} \partial_\theta
\left(\sin\theta\partial_\theta\Psi\right) \nonumber \\
&&
+\left[ {1 \over \sin^2\theta} -{a^2 \over\Delta} \right]
         \partial_{\phi\phi}\Psi 
+\, 2s \left[{a (r-M)\over \Delta} 
+  {i \cos\theta \over\sin^2\theta}
\right] 
\partial_\phi\Psi \nonumber \\
&& - \left(s^2 \cot^2\theta - s \right) \Psi = 0 
\label{teuk}  
\end{eqnarray}
The actual meaning of $\Psi$ for various spin-fields is rather complex, so we will only worry about two special cases here. Firstly, for $s=0$ $\Psi$ 
represents the scalar field ($\Phi$) 
itself and in the limit $a\to 0$ we recover
the Schwarzschild scalar wave equation.  For $s=\pm2$
the $\Psi$  corresponds to the Weyl curvature scalars $\Psi_0$ and $\Psi_4$ 
that directly represent the gravitational-wave degrees of freedom.

The great breakthrough that followed Teukolsky's derivation of 
(\ref{teuk}) was that one could now separate the variables also
for Kerr perturbations.
For our present purposes it is sufficient
to note that this essentially corresponds to assuming that i) the
time-dependence of the perturbation is accounted for via Fourier
transformation, and ii) there exists a suitable set of angular
function
that can be used to separate the coordinates $r$ and $\theta$. In the
case of scalar perturbations, the angular functions turn out to be 
standard spheroidal wavefunctions.      
Knowing this we assume a representation (for each given
integer $m$)
\begin{equation}
\Phi = \int d\omega\, e^{-i\omega t} \sum_{l=0}^\infty R_{lm}(r,\omega) 
S_{lm}(\theta,\omega) \ ,
\end{equation}
where it should be noted that the angular functions depend explicitly on the
frequency $\omega$. That is, they are intrinsically time-dependent
functions. After separation of variables, the problem is reduced to a
single ordinary differential equation for $R_{lm}(r,\omega)$. This
equation can be written as  
\begin{equation}
{d^2 R_{lm} \over dr_\ast^2} + 
\left[ {K^2 + (2am\omega -a^2\omega^2 - E)\Delta \over (r^2+a^2)^2}
-{dG \over dr_\ast} - G^2 \right] R_{lm} = 0 \ ,
\label{radeq}\end{equation}
where
$K = (r^2 + a^2)\omega - am$, 
$G = r\Delta/(r^2+a^2)^2$, and
the tortoise coordinate $r_\ast$ is defined from 
$ dr_\ast = (r^2 + a^2)/\Delta \, dr$.
The variable $E$ is the angular separation constant. 
In the limiting case
$a\to 0$, it reduces to $l(l+1)$, and for nonzero $a$ it can be obtained
from a power series in $a\omega$. It should be noted that 
$E$ is real valued for real frequencies.

\subsection{Quasinormal modes}

The physical solution to (\ref{radeq}) is defined by the asymptotic behaviour
(cf. the Schwarzschild result)
\begin{equation}
R_{lm} \sim \left\{ \begin{array}{ll}  e^{-i(\omega-m\,\omega_+)\,r_\ast} 
\quad &\mbox{as } r\to r_+ \ , \\
A_{\rm out} e^{i\omega r_\ast} + A_{\rm in} e^{-i\omega r_\ast} 
\quad &\mbox{as } r\to +\infty \ . \end{array}
\right.
\label{asymp}\end{equation}
where $ \omega_+ \equiv a /2\,M\,r_+$
is the angular velocity of the event horizon.

From this we see that we can  define
the quasinormal modes of a Kerr black hole in the same way 
as we did in the spherically symmetric case. Furthermore, these
modes can also be calculated using Leaver's continued fraction method.
The results can be summarised as follows:
Recall that in the
non-rotating case the modes occur in complex-frequency
pairs $\omega_q$ and $-\bar{\omega}_q$ (the bar denotes
complex conjugation). This is apparent in Figure~\ref{leaver}.
As the black hole spins up, each Schwarzschild mode splits into a multiplet 
of $2l+1$ distinct modes (in analogy with the Zeeman splitting in
quantum mechanics). These modes are associated with the various
values of $m$, where $-l \le m \le l$, 
which determine the modal dependence on the azimuthal
angle through $e^{im\varphi}$. As is straightforward to
deduce, modes for which $\mbox{Re } \omega_q$ and $m$ have the
same sign are co-rotating with the black hole. Similarly, modes
such that $\mbox{Re } \omega_q$ and $m$ have opposite
signs are counter-rotating. The effect that rotation has
on the mode-frequencies can, to some extent, be deduced from this fact.

Let us first consider the counter-rotating modes: These will appear
to be slowed down by inertial frame dragging close to the
black hole. Hence, their oscillation frequencies will tend
to decrease as $a\to M$. At the same time,  numerical
calculations show that the damping rate stays 
almost constant. For the co-rotating modes, the
effect is the opposite. Frame-dragging tends to increase
the frequencies. Additionally, the 
modes become much longer lived.
The available numerical results for co-rotating modes
are well approximated by (cf. \cite{leaver})
\begin{equation}
\mbox{Re } \omega_0 \approx {1\over M} \left[ 1 -
{63\over 100} (1-a/M)^{3/10} \right] \ ,
\end{equation}
and
\begin{equation}
\mbox{Im } \omega_0 = {(1-a/M)^{9/10} \over 4M }
 \left[ 1 -
{63\over 100} (1-a/M)^{3/10} \right] \ . 
\end{equation}
From this we can see that the mode becomes
undamped in the limit $a=M$. One can actually show 
(since the case $a= M$ is 
amenable to analytic methods) that there
exists an infinite sequence of real resonant 
frequencies with a common
limiting point in that case.
The limiting frequency is  $\omega = m/2M$, which we will later
 show to 
be the upper limit for so-called
super-radiance. That the modes become undamped
can be understood from the fact that the angular frequency of an
extreme
Kerr black hole is $1/2M$. As $a\to M$ the long lived
quasinormal modes essentially
rotate uniformly with the black hole, and as a consequence they
do not radiate strongly.      

\begin{figure}[h]
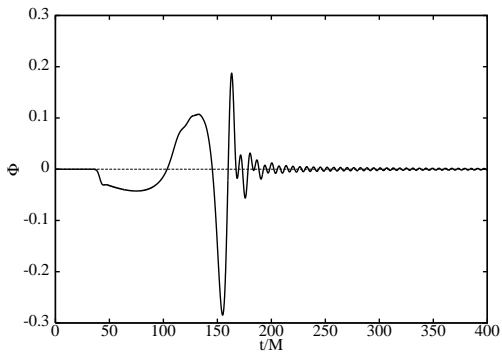

\caption{The response of a near extreme Kerr black hole after 
a Gaussian scalar wave pulse has impinged upon it. The main features
are the same as in the Schwarzschild case (Figure~\ref{vishu}), 
but here the quasinormal
mode ringing is much slower damped.  }
\label{teukev}
\hrule
\tkev
\hrule
\end{figure}

Provided that the modes of close to extreme rotating black
holes will be excited by some realistic astrophysical 
process, the fact that these modes can be very long-lived
would greatly improve the chances for detection with
future gravitational-wave detectors.   Hence, it is of
interest to investigate the excitation of these long lived modes. 
Recent work has shown that the modes tend to be harder
to excite than the short-lived Schwarzschild modes,
but that the slowly damped modes nevertheless dominate the 
 emerging signals. An example of this is shown in
Figure~\ref{teukev}.

\subsection{Superradiant scattering}   

Given the prescribed asymptotic behaviour (\ref{asymp}), 
together with that for the complex conjugate of $R_{lm}$ and the fact that
two linearly independent solutions to (\ref{radeq}) must lead
to a constant Wronskian, it is not difficult to show that
\begin{equation} 
( 1 - m\,\omega_+/\omega ) |T|^2  = 1 - |S|^2  \ .
\end{equation}
where we have introduced the transmission and reflection coefficients
as in the Schwarschild case;
\[
|T|^2 = \left| {1 \over A_{\rm in} } \right|^2 \ ,\qquad \qquad
|S|^2 = \left| {A_{\rm out}\over A_{\rm in}} \right|^2 \ ,
\label{refle}
\]

From this result, it is evident that the scattered waves are amplified
($|S|^2>1$) if
\begin{equation}
\omega < m\,\omega_+
\end{equation}
This is known as superradiance and it is the wave-analogue to the
Penrose process that we described earlier.
Its existence
implies that it would in principle be possible to mine a rotating
black hole for some of its rotational energy. 

In Figure~\ref{fig1}, we show a sample of results for the reflection
coefficient in the case when $l=m=2$. These results were obtained by 
a straightforward integration of (\ref{radeq}) and subsequent
extraction of ${\cal R}$. The maximum amplification in this case
is a minute 0.2 \%. 
Similar results for electromagnetic waves and 
gravitational perturbations show that
the maximum amplification is 0.3\% for scalar waves, 
4.4\% for electromagnetic waves and as large as
138\% for gravitational waves.

Given the results in Figure~\ref{fig1}, it is worth pointing out that
they agree with standard conclusions regarding the apparent ``size''
of a rotating black hole as seen by different observers.
A rotating black hole will
appear larger to a particle moving around it in a retrograde orbit than to a
particle in a prograde orbit. This is illustrated by the fact that the
 unstable circular
photon orbit (at $r=3M$ in the non-rotating case) is located at
$r=4M$ for a retrograde photon, while it lies at $r=M$ for a prograde
photon. The results in Figure~\ref{fig1} illustrate the same effect:
In our case, we have prograde motion when $\omega/m$ is positive and
retrograde motion when $\omega/m$ is negative. The data in
Figure~\ref{fig1}
correspond to $m=2$, and the enhanced reflection for positive
frequencies
as $a\to M$ has the 
effect that the black hole ``looks smaller'' to such waves.  
Conversely, the
slightly decreased reflection for negative frequencies leads to the
black hole appearing ``larger'' as $a\to M$. A sample of results 
showing this effect is in Table~\ref{t4.1}.

\begin{figure}[h]
\caption{\label{fig1} Reflection coefficient for different
values of $a$ in the case $l=m=2$. Upper panel: 
 As $a$ increases, there is
a clearly enhanced reflection of prograde waves ($\omega>0$ in the
figure)
while the overall reflection of retrograde waves ($\omega<0$ in the
figure) decreases somewhat. Lower panel: A blow-up of the result for prograde waves unveils a maximum amplification due to
superradiance of 0.187\%. }
\hrule
\srad
\hrule
\end{figure}

\begin{table}[h]
\caption{The apparent size $b$ of a Kerr black hole as viewed along 
the rotation axis. The values are all for $a=0.9M$ and are estimated
 from the total absorbtion cross section $\sigma^{\rm abs}\approx
\pi b^2$. Positive frequencies co-rotate with
the black hole whereas negative ones are counter-rotating. 
The values should be compared to $b=5.2M$ for a Schwarzschild black hole.}
\begin{center}
\begin{tabular}[t]{c|cccc} 
$\omega M$ & $-1.5$ & $-0.75$ & $0.75$ & $1.5$\\ \hline
$\sigma^{\rm abs}$ & $80.3M^2$ & $88.7M^2$ & $62.5M^2$ & $36.5M^2$ \\ \hline
$b$ & $5.06M$ & $5.31M$ & $4.46M$ & $3.41M$ \\ 
\end{tabular}
\end{center}
\label{t4.1}
\end{table}

\subsection{Scattering of waves by Kerr black holes}

One can analyse the scattering of monochromatic waves by a Kerr black hole
in much the same way as we approached the Schwarzschild case. 
The same is true also for wave fields other than scalar waves. 
The analysis of electromagnetic and gravitational wave scattering is,
in principle, identical to that for scalar waves. However, in the case of 
gravitational waves an additional complication enters: Gravitational
waves come with two different polarisations. This means that the 
scattering amplitude  consists of a sum
of their individual contributions, and that 
the cross section may show features due to 
interference between these two contributing terms.
This effect has not been explored in detail as yet.

Similarly, despite a few studies, the full details of scattering from
rotating black holes remain to be understood. 
Scattering of gravitational waves incident along the axis of symmetry 
of a Kerr black hole show that the scattering cross section depends 
in a complicated way on the rotation parameter $a$. 
One would essentially expect Kerr scattering to be different
because of two effects that do not
exist for nonrotating holes: superradiance and the 
polarization of the incident wave.
For incidence along
the symmetry axis of a rotating black hole 
one can have either co- or counter-rotating waves. 
The two cases  lead to quite different results. 
Although the general 
features of the corresponding cross sections are similar
they show different structure in the backward direction. This 
possibly arises  because of the phase-difference between
the two polarisations of gravitational waves.
As for superradiance, it has been suggested that 
it tends to  wash out interference minima. 
Further details can be found in the book by Futterman, Handler
and Matzner~\cite{futterman}.

\begin{acknowledgements}
Figure \ref{suivi2} is from unpublished work by BJ and Antoine
Folacci. We would like to thank P. L. Jensen for supplying the
original artwork for Figure \ref{rocket}.

\end{acknowledgements}

\end{document}